\documentclass[twoside,twocolumn]{article}

\usepackage{tabulary,graphicx,times,caption,fancyhdr,amsfonts,amssymb,amsbsy,latexsym,amsmath}
\usepackage[utf8]{inputenc}
\usepackage{url,multirow,morefloats,floatflt,cancel,tfrupee,textcomp,colortbl,xcolor,pifont}
\usepackage[nointegrals]{wasysym}
\usepackage{ifxetex}
\ifxetex\else
  \usepackage{dblfloatfix}
\fi

\def\URL#1#2{\@ifundefined{href}{#2}{\href{#1}{#2}}}

\makeatother

\usepackage[paperheight=11in,paperwidth=8.3in,margin=2.5cm,headsep=.7cm,top=2.5cm]{geometry}
\usepackage[T1]{fontenc}

\renewenvironment{abstract}
	{\trivlist\item[]\leftskip0pt\par\vskip4pt\noindent
  	\textbf{\abstractname}\mbox{\null}\\}
	{\par\noindent\endtrivlist}

\def\keywords#1{\par\medskip\par\noindent\textbf{Keywords}: #1\par}

 \date{} \emergencystretch 8pt

\makeatletter
\def\author#1{\gdef\@author{\hskip-\tabcolsep%
	\parbox{\textwidth}{\raggedright\bfseries#1\\[1pc]}}}
\def\address[#1]#2{\g@addto@macro\@author{\\\hskip-\tabcolsep\parbox{\textwidth}{\raggedright%
	\normalsize\normalfont\textsuperscript{#1}#2}}}
\let\addresslink\textsuperscript
\def\correspondence#1{\g@addto@macro\@author{\\\hskip-\tabcolsep\parbox{\textwidth}{\raggedright%
	\vspace*{10pt}\normalsize\normalfont~\\#1~\\[12pt]}}}
\def\email#1{\g@addto@macro\@author{\\\hskip-\tabcolsep\parbox{\textwidth}{\raggedright%
	\normalsize\normalfont Emails: #1}}}

\def\title#1{\gdef\@title{\vspace*{-30pt}%
	\raggedright\textbf{\@journaltitle}~\\%
  \raggedright\bfseries\ifx\@articleType\@empty\vspace*{20pt}\else%
  \vspace*{20pt}\@articleType\vspace*{20pt}\\\fi#1}}
\let\@journaltitle\@empty \def\journaltitle#1{\gdef\@journaltitle{{\normalfont\itshape#1}}}
\let\@articleType\@empty \def\articletype#1{\gdef\@articleType{{\normalfont\itshape#1}}}

\let\@runningHead\@empty \def\RunningHead#1{\gdef\@runningHead{{\normalfont #1}}}

\usepackage{fancyhdr}
\fancypagestyle{headings}{\fancyhf{}
  \fancyhead[R]{\itshape\@runningHead}
  \fancyfoot[C]{\thepage}}
\fancypagestyle{plain}{%
	\fancyhf{}\fancyhead[R]{}
  \fancyfoot[C]{\thepage}}
\makeatother

\usepackage{float,xcolor}

\usepackage{amsmath}
\usepackage{amssymb}
\usepackage[linesnumbered,ruled,vlined]{algorithm2e} 

\usepackage[utf8]{inputenc} 
\usepackage[T1]{fontenc}    
\usepackage{hyperref}       
\usepackage{url}            
\usepackage{booktabs}       
\usepackage{amsfonts}       
\usepackage{nicefrac}       
\usepackage{microtype}      
\usepackage{cleveref}       
\usepackage{lipsum}         
\usepackage{graphicx}
\usepackage{doi}
\usepackage{tikz}
\usetikzlibrary{automata,positioning,arrows.meta,calc,shapes, shapes.geometric}
\usepackage{pgfplots}
\pgfplotsset{compat=newest}
\usetikzlibrary{pgfplots.polar}

\usepackage{multirow}
\usepackage{longtable}
\usepackage{float}

\tikzstyle{startstop} = [rectangle, rounded corners, minimum width=3cm, minimum height=1cm,text centered, draw=black, fill=red!30]
\tikzstyle{process} = [rectangle, minimum width=3cm, minimum height=1cm, text centered, draw=black, fill=blue!30]
\tikzstyle{decision} = [diamond, minimum width=3cm, minimum height=1cm, text centered, draw=black, fill=green!30]
\tikzstyle{arrow} = [thick,->,>=stealth]

\newcolumntype{P}[1]{>{\centering\arraybackslash}p{#1}}

\usepackage[
  style=numeric-comp,
  datamodel=software, 
  abbreviate=false,
  sorting=none,
  backend=bibtex,
  bibencoding=utf8,
  url=true,
  doi=true,
  defernumbers,
  maxcitenames=3,
  defernumbers=false,
  maxbibnames=100]{biblatex}
\bibliography{references}

\definecolor{myColor}{rgb}{0,0.1,0.55} 
\journaltitle{Journal of Computer Networks and Communications}
\articletype{Research Article} 

\begin{document}

\title{Optimizing the 4G--5G Migration: A Simulation-Driven Roadmap for Emerging Markets}

\author{%
		Desire~Guel \addresslink{1},
  	Justin~Pegd-Windé~Kouraogo \addresslink{1} and
  	 Kouka~Kouakou~Nakoulma \addresslink{1,2}
    }
		
\address[1]{Department of Computer Science, Joseph KI-ZERBO University, Ouagadougou, Burkina Faso}
\address[2]{Autorité de Régulation  des Communications, Electroniques et des Postes (ARCEP), Ouagadougou, Burkina Faso }

\correspondence{Correspondence should be addressed to 
    	Desire~Guel: desire.guel@ujkz.bf}

\email{desire.guel@ujkz.bf (Desire~Guel), kouraogo@gmail.com (Justin~Pegd-Windé~Kouraog), nakkouka@gmail.com (Kouka~Kouakou~Nakoulma)}%

\RunningHead{Optimizing the 4G--5G Migration}

\maketitle 

\begin{abstract}
Deploying fifth–generation (5G) networks in emerging markets requires strategies that balance ambitious performance targets with fiscal, spectrum and infrastructure constraints. We develop a simulation–driven assessment using \textsc{MATLAB} to quantify how key radio and architectural levers—multiple–input multiple–output (MIMO) modes (beamforming, diversity, spatial multiplexing), intra-/inter–band carrier aggregation (CA), targeted spectrum refarming to New Radio (NR), millimeter–wave (mmWave) propagation with blockage/rain and Non–Standalone (NSA) versus Standalone (SA) core options—shape capacity, coverage, latency and interference robustness. We also evaluate Device–to–Device (D2D) and Machine–to–Machine (M2M) communication as complements to wide–area access.

Beamforming improves edge SNR by $\sim$3–6\,dB and stabilizes links at low SNR. Spatial multiplexing dominates at moderate/high SNR, increasing spectral efficiency via multi–stream gains. Aggregate throughput scales strongly with CA: moving from 1 to 5 component carriers ($20$\,MHz each) raises peak rates from $\sim$200\,Mbps to $\sim$1\,Gbps at 30\,dB SNR; water–filling adds 5–12\% over equal–power at mid–SNR. Targeted refarming of mid–band to NR increases median urban throughput by 60–90\% (rural +40–70\%) while preserving coverage by retaining sub–1\,GHz layers. At 28\,GHz, rain and human blockage add $\sim$8–30\,dB excess loss; viable deployments concentrate in LOS hot–zones with narrow–beam arrays, requiring ISD $\approx$140–170\,m to meet 100\,Mbps with 100\,MHz. NSA first delivers broader initial coverage (about 78–82\%) than SA on the same grid (about 64–70\%) by reusing LTE/EPC; SA becomes attractive as transport ($\ge$10\,Gb/s, sub–5\,ms RTT) and site density mature. D2D achieves $\sim$1\,ms one–way latency for proximity services and grant–efficient M2M sustains throughput as device density increases.

We synthesize these findings into a decision–oriented roadmap for resource–constrained settings: anchor early NR on NSA, use CA–centric spectrum strategies with focused refarming, densify selectively in demand hotspots and follow a disciplined glide path to SA as transport and device ecosystems mature. The roadmap provides actionable guidance for operators and policymakers pursuing inclusive, cost–aware 5G migration.

\keywords{5G migration; emerging markets; MIMO; carrier aggregation; spectrum refarming; mmWave; NSA/SA; D2D; M2M}
\end{abstract}

\section{Introduction}
\label{sec:introduction}

The fifth generation (5G) of mobile networks promises transformative improvements in data throughput, latency and device density, enabling new verticals such as industrial IoT and autonomous systems \cite{liyanage_comprehensive_2018,vannithamby_towards_2017}. However, realizing these benefits in emerging markets remains highly context-sensitive. In countries like Burkina Faso, operators face tight capital constraints, legacy-heavy device ecosystems, fragmented backhaul infrastructure and pronounced urban–rural disparities. As of 2023, Burkina Faso counted 23 million mobile subscriptions, with over 85\% 4G population coverage but uneven distribution across regions and operators \cite{ARCEP_2021}. Spectrum holdings are modest: sub-1 GHz and 1.8/2.1 GHz allocations dominate, with only recent auctions extending into the 2.6 GHz range. These constraints complicate 5G deployment decisions—particularly between non-standalone (NSA) and standalone (SA) architectures—and call for context-aware technical and policy analysis \cite{prasad_5g_2014,chandramouli_5g_2019}.

NSA deployments, which anchor 5G New Radio (NR) on existing LTE/EPC infrastructure, offer a rapid and cost-effective migration path while ensuring service continuity. By contrast, SA deployments introduce a new 5G core (5GC), enabling advanced functionalities such as network slicing and ultra-reliable low-latency communication (URLLC), but at the expense of higher transport and site-density requirements \cite{Huawei2020,Penttinen2019}. Choosing between NSA and SA is not merely a technical decision—it is tightly coupled with spectrum availability, backhaul readiness and terminal diversity. It also interacts with radio-layer options such as MIMO configuration, carrier aggregation (CA), mid-band refarming and the potential use of mmWave frequencies \cite{vannithamby_5g_2020,manner_spectrum_2021}.

This study develops a simulation-driven framework tailored to the constraints of emerging markets to evaluate how various 5G radio and core network configurations impact performance and deployment viability. We investigate how MIMO modes trade off spectral efficiency and robustness under realistic SNR conditions; quantify CA gains under spectrum fragmentation; assess the impact of low-/mid-band refarming on urban/rural coverage; evaluate mmWave feasibility under diffraction and blockage scenarios; and compare NSA-first versus early-SA strategies in terms of coverage, time-to-benefit and transport readiness. These five research questions are addressed respectively via: (i) link-level simulation of MIMO/CA; (ii) refarming scenario analysis; (iii) mmWave path-loss modeling; and (iv) NSA/SA deployment experiments under capacity and latency constraints.

Our main contributions are fourfold. First, we propose a context-aware simulation environment that couples canonical information-theoretic models with operational constraints representative of low-/middle-income settings. Second, we provide a comparative analysis of radio configurations—MIMO, CA, mmWave—under spectrum and infrastructure limitations. Third, we derive a deployment roadmap linking NSA-first strategies to targeted spectrum refarming and densification, identifying transport upgrade thresholds critical for SA adoption. Finally, we draw regulatory and investment implications for markets with shared constraints, highlighting the role of flexible licensing, infrastructure sharing and staggered rollout to reduce capital intensity. The remainder of this paper is organized as follows: Section~\ref{sec:BackgroundAndRelatedWork} reviews the relevant literature and deployment strategies; Section~\ref{sec:methodology} formalizes the models and simulation approach; Section~\ref{sec:results} presents results and discussion; and the paper concludes with recommendations and future research directions.

\section{Theoretical Framework for 5G Deployment Optimization}
\label{sec:theory}
This section establishes a concise mathematical basis for the main levers of 5G deployment in emerging markets—MIMO processing, CA, spectrum allocation andmmWave propagation—so as to relate design parameters to network-level performance. Unless stated otherwise, bold lowercase/uppercase denote vectors/matrices, $(\cdot)^H$ the Hermitian and$\log$ is base-2.

\subsection{System Model}
\label{subsec:system_model}

We consider a narrowband, block-fading MIMO baseband link with \( N_t \) transmit and \( N_r \) receive antennas. The received signal is modeled as:
\begin{equation}
\label{eq:rxmodel}
\mathbf{y} = \mathbf{H}\mathbf{x} + \mathbf{n},
\end{equation}
where \( \mathbf{H} \in \mathbb{C}^{N_r \times N_t} \) is the complex channel matrix, \( \mathbf{x} \in \mathbb{C}^{N_t} \) is the transmitted signal with average power constraint \( \mathbb{E}\{\lVert\mathbf{x}\rVert^2\} = P_t \) and \( \mathbf{n} \sim \mathcal{CN}(\mathbf{0}, \sigma_n^2 \mathbf{I}_{N_r}) \) is circular complex additive white Gaussian noise (AWGN). The noise power per receive antenna is \( \sigma_n^2 = N_0 B \), where \( B \) is the channel bandwidth and \( N_0 \) is the one-sided noise spectral density.

We assume a flat (narrowband) fading model with perfect channel state information at the receiver (CSI-R) and optional CSI at the transmitter (CSI-T), which is relevant in capacity computations involving water-filling. Each block-fading realization is considered independent across coherence intervals. The received signal-to-noise ratio (SNR) per antenna is defined as:
\begin{equation}
\label{eq:snr}
\mathrm{\rho \triangleq \frac{P_t}{N_t \sigma_n^2} = \frac{P_t}{N_t N_0 B}}.
\end{equation}

In engineering terms, the SNR in decibels is given by \( \rho_{\mathrm{dB}} = 10\log_{10}(\rho) \) and a practical link budget must account for implementation losses, typically in the range of 1–2 dB (e.g., coding or synchronization penalties). The parameters \( N_t \), \( N_r \), \( B \) and \( \rho \) will serve as primary inputs to the simulation scenarios in Section~\ref{sec:methodology}.

\subsection{MIMO Capacity Analysis}
\label{subsec:mimo_capacity}
With perfect CSI at the receiver and equal power across transmit antennas, the spectral efficiency (bits/s/Hz) is defined as per \cite{Foschini1998}:
\begin{equation}
\label{eq:mimo_ce}
\eta \;=\; \log \det\!\Big(\mathbf{I}_{N_r}+\rho\,\mathbf{H}\mathbf{H}^H\Big),
\end{equation}
and the Shannon capacity is $C = B\,\eta$ \cite{Cover2006,Shannon1948}. Special cases:
\begin{itemize}
  \item \textbf{SISO} ($N_t{=}N_r{=}1$): $\displaystyle \eta=\log\!\big(1+\tfrac{P_t}{\sigma_n^2}|h|^2\big)$ \cite{Cover2006,Shannon1948}.
  \item \textbf{SIMO} ($N_t{=}1$, $N_r{>}1$): $\displaystyle \eta=\log\!\big(1+\tfrac{P_t}{\sigma_n^2}\lVert\mathbf{h}\rVert^2\big)$ \cite{Mo2014}.
  \item \textbf{MISO} ($N_t{>}1$, $N_r{=}1$): with transmit beamforming, $\displaystyle \eta=\log\!\big(1+\tfrac{P_t}{\sigma_n^2}\lVert\mathbf{h}\rVert^2\big)$ \cite{skold_4g_2011}.
  \item \textbf{Rank-$r$ MIMO}: letting $\lambda_1,\dots,\lambda_r$ be the nonzero eigenvalues of $\mathbf{H}\mathbf{H}^H$,
  \begin{equation}
  \label{eq:rank_def}
  \eta=\sum_{i=1}^{r}\log\!\Big(1+\rho\,\lambda_i\Big).
  \end{equation}
\end{itemize}
If CSI is also available at the transmitter, water-filling across the eigenmodes with powers $p_i$ maximizes $\eta$ subject to $\sum_i p_i{=}P_t$:
\begin{equation}
\label{eq:waterfilling_mimo}
\mathrm{p_i=\Big[\mu-\frac{\sigma_n^2}{\lambda_i}\Big]^+,\qquad \sum_{i=1}^r p_i=P_t}, \quad \text{\cite{Cover2006}}.
\end{equation}
These expressions formalize how (i) more antennas, (ii) higher SNR and(iii) better channel rank/conditioning increase throughput \cite{skold_4g_2011,vannithamby_5g_2020}.

\subsection{Carrier Aggregation and Bandwidth Optimization}
\label{subsec:ca}

Carrier Aggregation (CA) enables simultaneous transmission over \( N_c \) component carriers (CCs), each with bandwidth \( B_i \), received power \( P_i \) and channel gain \( |h_i|^2 \). Assuming each CC experiences flat AWGN conditions, the aggregate downlink throughput is given by:
\begin{equation}
\label{eq:ca_capacity}
\mathrm{T = \sum_{i=1}^{N_c} B_i\,\log_2\!\left(1+\gamma_i\right)}, \quad 
\mathrm{\gamma_i \triangleq \frac{P_i |h_i|^2}{N_0 B_i}},
\end{equation}
where \( \gamma_i \) is the SNR per carrier and \( \sum_i P_i \le P_t \) is the total transmit power budget \cite{Shannon1948,Cover2006}.

To maximize \( T \), optimal power allocation across CCs follows scalar water-filling over the effective gain-to-noise ratios \( \mathrm{g_i \triangleq |h_i|^2 / (N_0 B_i)} \):
\begin{equation}
\label{eq:waterfilling_ca}
\mathrm{P_i = \left[\mu - \frac{1}{g_i}\right]^+}, \quad \mathrm{\text{with } \sum_{i=1}^{N_c} P_i = P_t},
\end{equation}
where \( \mu \) is a water level satisfying the total power constraint and \( [\cdot]^+ \) denotes the projection onto non-negative values.

\begin{algorithm}[h]
\small
\DontPrintSemicolon
\caption{Water-Filling Power Allocation for Carrier Aggregation}
\label{alg:ca_waterfilling}
\KwIn{\( \{g_i\}_{i=1}^{N_c} \): effective gains, \( P_t \): total power}
\KwOut{\( \{P_i\}_{i=1}^{N_c} \): allocated powers}
Sort CCs such that \( g_{(1)} \ge \dots \ge g_{(N_c)} \); set \( k \leftarrow 1 \). \;
\While{\( k \le N_c \)}{
Compute tentative level \( \mu = \dfrac{P_t + \sum_{j=1}^{k} \tfrac{1}{g_{(j)}}}{k} \). \;
Set \( P_{(j)} \leftarrow \left[\mu - \tfrac{1}{g_{(j)}}\right]^+ \) for \( j = 1,\dots,k \). \;
\If{\( P_{(k)} \le 0 \)}{Set \( k \leftarrow k{-}1 \); recompute \( \mu \); \textbf{break}.}
\Else{Set \( k \leftarrow k{+}1 \).}
}
\Return \( \{P_i\} \). \;
\end{algorithm}

\vspace{0.5em}
\noindent Consider \( N_c = 3 \) carriers, each of bandwidth \( B_i = 100 \)~MHz and a total transmit power \( P_t = 1 \)~W. Let channel gains be \( |h_1|^2 = 1.2 \), \( |h_2|^2 = 0.9 \), \( |h_3|^2 = 0.5 \) and assume \( N_0 = 10^{-20} \)~W/Hz. Then, effective gain-to-noise ratios are:
\[
\mathrm{g_i = \frac{|h_i|^2}{N_0 B_i}} \quad \Rightarrow \quad 
\begin{cases}
\mathrm{g_1 = \frac{1.2}{10^{-20} \cdot 10^8} = 1.2 \times 10^{12}} \\
\mathrm{g_2 = 0.9 \times 10^{12}}, \quad
\mathrm{g_3 = 0.5 \times 10^{12}}
\end{cases}
\]
Applying Algorithm~\ref{alg:ca_waterfilling}, we obtain power allocations \( \mathrm{P_1 = 0.41} \)~W, \( \mathrm{P_2 = 0.36} \)~W and \( \mathrm{P_3 = 0.23} \)~W. These values yield respective per-CC throughputs via~\eqref{eq:ca_capacity}, leading to a total throughput \(\mathrm{ T \approx 1.93} \) Gbps.

\vspace{0.5em}
\noindent In multi-user settings, this formulation can be extended per user and carrier, subject to scheduling policies such as round-robin or proportional-fair allocation. The scheduler assigns \( \mathrm{P_i^{(u)}} \) per user \( u \), while enforcing \( \mathrm{\sum_{u,i} P_i^{(u)} \le P_t} \). Scheduling decisions then balance fairness, throughput and radio conditions.

\subsection{Spectrum Allocation Optimization}
\label{subsec:spectrum}
Let $x_i$ denote the spectrum portion (Hz) assigned to entity $i$ (UE, slice, or service) with utility $\mathrm{U_i(x_i)}$ (e.g., achievable capacity under a given PHY/MAC). A generic allocation reads
\begin{equation}
\label{eq:opt_spectrum}
\mathrm{\max_{\mathbf{x}\succeq \mathbf{0}}\;\sum_{i=1}^{N} U_i(x_i)\quad
\text{s.t.}\quad \sum_{i=1}^{N} x_i \le S_{\text{total}}},
\end{equation}
where concave $U_i(\cdot)$ yields a convex program and KKT conditions lead to water-filling-like solutions. This captures policy choices (eMBB/URLLC/mMTC) and refarming decisions that are central in emerging markets \cite{manner_spectrum_2021}.

\subsection{mmWave Propagation and Path Loss}
\label{subsec:mmwave}

Millimeter-wave (mmWave) frequencies offer vast bandwidths for 5G but suffer from severe propagation losses. The free-space path loss (FSPL) increases with both frequency and distance. Using the Friis transmission formula, the loss at distance \( d \) and carrier frequency \( f \) is:
\begin{equation}
\label{eq:friis}
\mathrm{PL}_{\mathrm{FS}}(d,f) = 20\log_{10}\left( \frac{4\pi f d}{c} \right) \;\text{[dB]},
\end{equation}
where \( c \) is the speed of light.

To better reflect real-world propagation, 3GPP TR 38.901 recommends a log-distance model incorporating shadow fading:
\begin{equation}
\label{eq:logdist}
\begin{split}
\mathrm{PL}(d) &= \mathrm{PL}(d_0) + 10n \log_{10}\left( \frac{d}{d_0} \right) + X_\sigma, \\
\mathrm{X_\sigma} &\mathrm{\sim \mathcal{N}(0, \sigma^2)},
\end{split}
\end{equation}
where \( \mathrm{d_0} \) is a reference distance (typically 1 m), \( \mathrm{n} \) is the path-loss exponent and \( \sigma \) is the standard deviation of shadowing. For outdoor urban microcell (UMi) environments at 28 GHz, 3GPP specifies \( n_{\text{LOS}} = 2.1 \), \( \sigma_{\text{LOS}} = 3.5 \)~dB, while for NLOS, \( n_{\text{NLOS}} = 3.3 \), \( \sigma_{\text{NLOS}} = 7.0 \)~dB \cite{3GPP38901}.

Rain attenuation is significant at mmWave and modeled as:
\begin{equation}
\mathrm{A_r(d) = \gamma_r \cdot d }\quad \text{[dB]}, \quad \text{with } \mathrm{\gamma_r = k R^\alpha},
\end{equation}
where \( R \) is the rainfall rate (mm/h) and \( k, \alpha \) are frequency- and polarization-dependent constants from ITU-R P.838-3 \cite{UITR2017}. For Burkina Faso’s rainy season (e.g., \( \mathrm{R = 80} \) mm/h at 28 GHz, horizontal polarization), typical parameters are \( \mathrm{k = 0.15} \), \( \mathrm{\alpha = 0.9} \), yielding \(\mathrm{ \gamma_r \approx 3.3 }\) dB/km \cite{Mosleh2022}.

Obstacles introduce diffraction losses. For a single knife-edge obstacle of height \( h \), the excess loss is modeled using the Fresnel–Kirchhoff parameter \( \nu \), as:
\begin{equation}
\label{eq:knifeedge}
\mathrm{L_{\text{diff}} = 20\log_{10}(\nu), \quad 
\nu = \frac{h \sqrt{2(d_1 + d_2)}}{\sqrt{\lambda d_1 d_2}}},
\end{equation}
where \( d_1, d_2 \) are distances from the transmitter and receiver to the obstacle and \( \lambda \) is the signal wavelength.

The total propagation loss at mmWave thus combines these effects:
\begin{equation}
\label{eq:total_loss}
\mathrm{PL}_{\text{total}}(d) = \mathrm{PL}(d) + A_r(d) + L_{\text{diff}} \quad \text{[dB]}.
\end{equation}

\noindent These propagation characteristics justify the use of highly directional beamforming, adaptive blockage-aware scheduling and dense base station deployments in mmWave networks \cite{rodriguez_fundamentals_2015,Mosleh2022,3GPP38901}.

\subsection{Low-Latency Communication (D2D and M2M)}
\label{subsec:d2d_m2m}
D2D bypasses the base station for proximity services, while M2M supports wide-area device traffic. A minimal latency budget for a single hop can be decomposed as
\begin{equation}
\label{eq:latency_generic}
\mathrm{L \;=\; L_{\text{proc}}+L_{\text{queue}}+L_{\text{MAC}}+L_{\text{prop}}},
\end{equation}
where $\mathrm{L_{\text{prop}}\!=\!d/c}$ is the propagation delay. Under an idealized MAC (negligible processing/queuing), the D2D one-shot exchange over a periodic Tx/Rx cycle of frequency $f_{\text{TxRx}}$ yields
\begin{equation}
\label{eq:d2d_latency}
\mathrm{L_{\text{D2D}}\approx \frac{1}{f_{\text{TxRx}}}},
\end{equation}
whereas a wide-area M2M hop of length $d$ adds propagation delay:
\begin{equation}
\label{eq:m2m_latency}
\mathrm{L_{\text{M2M}}\approx \frac{d}{c}+\frac{1}{f_{\text{TxRx}}}},\quad \text{\cite{Gandotra2017,Verma2016}}.
\end{equation}
In practice, $\mathrm{L_{\text{proc}}}$, $\mathrm{L_{\text{queue}}}$ andgrant/scheduling effects dominate the budget; nonetheless \eqref{eq:d2d_latency}–\eqref{eq:m2m_latency} highlight the inherent latency advantage of D2D for URLLC-type services.

\begin{table*}[h!]
\centering
\caption{Typical Specific Rain Attenuation $\gamma_r$ (dB/km) for Horizontal Polarization \cite{Mosleh2022,UITR2017}}
\label{tab:rain_attenuation}
\small
\begin{tabular}{|P{0.3\textwidth}|P{0.3\textwidth}|P{0.15\textwidth}|P{0.15\textwidth}|}
\hline
\textbf{Frequency (GHz)} & \textbf{Rain Rate (mm/h)} & \textbf{$k$} & \textbf{$\gamma_r=k R^\alpha$} \\ \hline
28 & 25 & 0.12 & 5.3 \\ \hline
38 & 25 & 0.17 & 7.6 \\ \hline
60 & 25 & 0.36 & 15.2 \\ \hline
73 & 25 & 0.48 & 20.4 \\ \hline
\end{tabular}
\end{table*}

\section{Background and Related Work}
\label{sec:BackgroundAndRelatedWork}
This section synthesizes foundational concepts and prior work—spanning the evolution from LTE/LTE-A to 5G NR, NSA versus SA deployment models, spectrum management, massive MIMO and carrier aggregation advances, mmWave propagation and D2D/M2M enablers—to situate our study within current standards and the state of the art.

\subsection{Evolution from 4G to 5G}
The evolution from LTE/LTE-Advanced to 5G has been widely documented with LTE-A features such as CA and advanced MIMO laying the technical groundwork for 5G New Radio (NR) \cite{perez_lte_2015,skold_4g_2011}. 5G extends these capabilities with a new air interface, expanded spectrum (including mmWave) anda cloud-native core to meet heterogeneous requirements across enhanced mobile broadband (eMBB), ultra-reliable low-latency communication (URLLC) andmassive machine-type communication (mMTC) \cite{liyanage_comprehensive_2018,rodriguez_fundamentals_2015,vannithamby_towards_2017,chandramouli_5g_2019}.

\begin{figure}[t]
\centering
\begin{tikzpicture}[node distance=.85cm and 2.5cm, on grid, auto]
\tikzstyle{block} = [rectangle, draw=blue!70, fill=blue!6, text width=7cm, text centered, rounded corners, minimum height=0.25cm]
\tikzstyle{arrow} = [thick,->,>=stealth,gray!90]

\node [block] (lte) {4G LTE (eMBB)};
\node [block, below=of lte] (lteA) {LTE-Advanced (Carrier Aggregation, MIMO)};
\node [block, below=of lteA] (transition) {5G Non-Standalone};
\node [block, below=of transition] (standalone) {5G Standalone (SBA, URLLC, mMTC)};
\node [block, below=of standalone] (future) {6G (THz, AI-native)};

\draw [arrow] (lte) -- (lteA);
\draw [arrow] (lteA) -- (transition);
\draw [arrow] (transition) -- (standalone);
\draw [arrow] (standalone) -- (future);
\end{tikzpicture}
\caption{Evolution path from LTE/LTE-A to 5G and beyond; see \cite{perez_lte_2015,skold_4g_2011,rodriguez_fundamentals_2015,vannithamby_towards_2017}.}
\label{fig:TransitionFrom4gTo5g}
\end{figure}
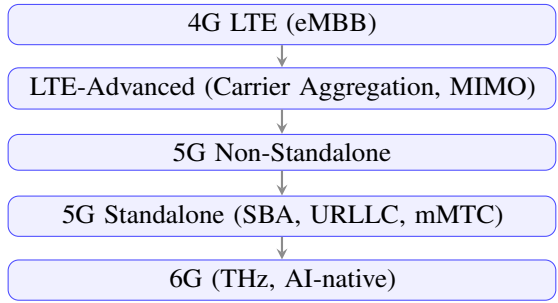

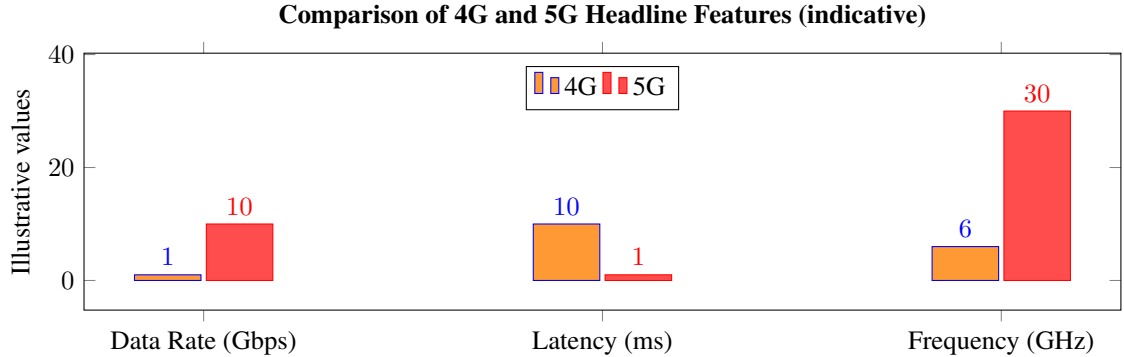
\begin{figure*}[t]
\centering
\begin{tikzpicture}
\begin{axis}[
    ybar,
    bar width=25pt,
    width=0.95\textwidth,
    height=5cm,
    enlargelimits=0.15,
    ylabel={Illustrative values},
    symbolic x coords={Data Rate (Gbps), Latency (ms), Frequency (GHz)},
    xtick=data,
    nodes near coords,
    nodes near coords align={vertical},
    ymin=0, ymax=35,
    legend style={at={(0.5,0.95)}, anchor=north, legend columns=2},
    title={Comparison of 4G and 5G Headline Features (indicative)},
    title style={font=\bfseries}
]
\addplot+[style={fill=orange!80}] coordinates {
    (Data Rate (Gbps), 1)
    (Latency (ms), 10)
    (Frequency (GHz), 6)
};
\addplot+[style={fill=red!70}] coordinates {
    (Data Rate (Gbps), 10)
    (Latency (ms), 1)
    (Frequency (GHz), 30)
};
\legend{4G, 5G}
\end{axis}
\end{tikzpicture}
\caption{Comparison of headline capabilities.}
\label{fig:4g5g_features_comparison}
\end{figure*}

5G targets diverse verticals by combining mmWave access, network slicing anda service-based core architecture (SBA) \cite{liyanage_comprehensive_2018,rodriguez_fundamentals_2015}. While eMBB drives peak-rate improvements, URLLC and mMTC impose stringent latency and connection-density requirements, respectively \cite{vannithamby_towards_2017,chandramouli_5g_2019}. Ongoing challenges include spectrum policy, energy efficiency andcost-effective densification \cite{prasad_5g_2014,manner_spectrum_2021}.

\begin{table*}[htbp]
\centering
\caption{Key Differences Between 4G and 5G (high-level)}
\label{tab:4g_5g_comparison}
\small
\begin{tabular}{|p{0.35\textwidth}|P{0.2\textwidth}|P{0.35\textwidth}|}
\hline
\textbf{Features} & \textbf{4G (LTE)} & \textbf{5G (NR)} \\ \hline
Headline data rate & Up to $\sim$1 Gbps & Multi-Gbps (deployment-dependent) \\ \hline
User-plane latency & $\sim$10 ms & $<\!1$ ms (URLLC target) \\ \hline
Operating bands & $<\!6$ GHz & Sub-6 GHz and $>\!24$ GHz (mmWave) \\ \hline
Core network & EPC & Service-based architecture (SBA) \\ \hline
Canonical use cases & eMBB & eMBB, URLLC, mMTC \\ \hline
\end{tabular}
\end{table*}

\subsection{Deployment Scenarios: NSA vs.\ SA}
Early 5G rollouts frequently adopt Non-Standalone (NSA); it reuse LTE/EPC for control-plane anchoring and enabling rapid time-to-market and broader coverage with limited capital outlay \cite{Huawei2020,GSMA2020}. Standalone (SA) introduces a native 5G core, unlocking slicing, low-latency user plane andscalable service exposure \cite{Penttinen2019}. For emerging markets, cost and backhaul constraints often favor an NSA-first path with progressive SA introduction.

\begin{figure*}[t]
\centering
\begin{tikzpicture}[node distance=2cm and 3cm, font=\footnotesize, every node/.style={align=center}]
\node[draw=blue!90, rectangle, text width=3cm, fill=blue!10] (ltecore) {4G Core (EPC)};
\node[draw, rectangle, text width=3cm, below left=of ltecore, xshift = 2cm, fill=green!10] (ltebs) {4G eNodeB};
\node[draw=orange!90, rectangle, text width=3cm, below right=of ltecore, xshift = -2cm, fill=orange!10] (5gnr) {5G NR (gNodeB)};
\node[draw=blue!90, rectangle, text width=3cm, right=5cm of ltecore, fill=blue!10] (5gcore) {5G Core (SBA)};
\node[draw=orange!90, rectangle, text width=3cm, below=of 5gcore, fill=orange!10] (5gsa) {5G gNodeB};

\draw[->, thick, blue!90] (ltebs) -- (ltecore);
\draw[->, thick, blue!90] (5gnr) -- (ltecore);
\draw[->, thick, blue!90] (5gsa) -- (5gcore);

\node[above=0.3cm of ltecore, font=\small\bfseries] {NSA Scenario};
\node[above=0.3cm of 5gcore, font=\small\bfseries] {SA Scenario};

\draw[decorate,decoration={brace,amplitude=10pt,mirror},xshift=0pt,yshift=-5pt]
  (ltebs.south west) -- (5gnr.south east) node [black,midway,yshift=-0.8cm] {\small NSA Deployment};

\draw[decorate,decoration={brace,amplitude=10pt,mirror},xshift=0pt,yshift=-5pt]
  (5gsa.south west) -- (5gsa.south east) node [black,midway,yshift=-0.8cm] {\small SA Deployment};
\end{tikzpicture}
\caption{NSA leverages LTE/EPC for accelerated rollout; SA enables full 5G core capabilities \cite{Huawei2020,GSMA2020,Penttinen2019}.}
\end{figure*}
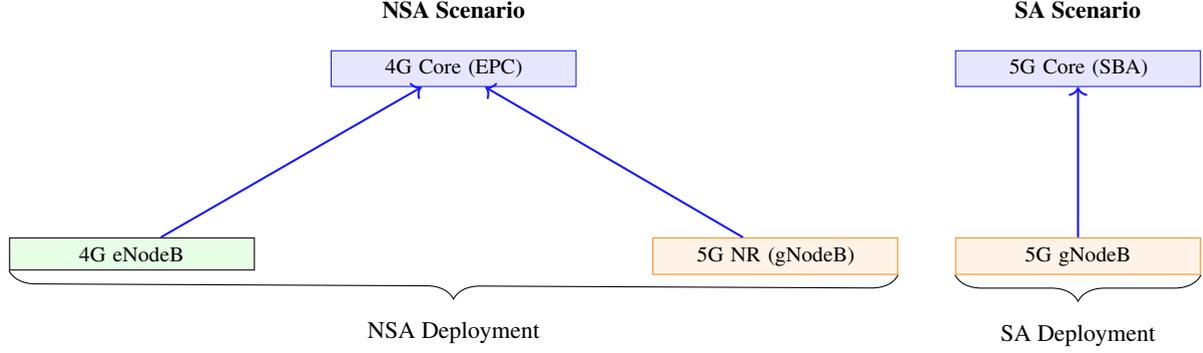

\subsection{Spectrum Management Challenges}
Efficient spectrum policy is pivotal to 5G outcomes, particularly where capital and fiber backhaul are constrained. Strategies include refarming LTE spectrum to NR, dynamic sharing andCA across low-, mid- andhigh-band assets \cite{Manner2021,ARCEP2021,UITR2017}. Table~\ref{tab:spectrum_allocation} contrasts typical 4G/5G usage patterns in developed versus emerging markets.

\begin{table*}[htbp]
\centering
\caption{Comparison of Spectrum Allocations in 4G and 5G.}
\label{tab:spectrum_allocation}
\small
\begin{tabular}{|p{1.75cm}|p{2.5cm}|p{3.0cm}|p{3.5cm}|p{3.5cm}|}
\hline
\textbf{Spectrum Range} & \textbf{4G Usage} & \textbf{5G Usage} & \textbf{Emerging Markets} & \textbf{Developed Markets} \\ \hline
Sub-1 GHz & Rural coverage & Extended rural coverage & Coverage focus; fewer small cells & Coverage with IoT integration \\ \hline
1--6 GHz & Mobile broadband & Enhanced mobile broadband & Primary for urban rollout & Dense urban/suburban coverage \\ \hline
$>\!24$ GHz & Not utilized & High-capacity hotspots & Limited adoption\textsuperscript{(*)} & Urban hotspots and venues \\ \hline
Cell types & Macro cells & Macro, micro, pico & Emphasis on macro/micro & All types incl.\ femto \\ \hline
Spectrum aggregation & Limited & Extensive CA & Growing adoption & Widely deployed \\ \hline
Use-case focus & Voice, basic Internet & IoT, AR/VR, smart cities & Limited IoT & Advanced IoT, industry \\ \hline
\end{tabular}
\begin{flushleft}
\textsuperscript{(*)}\;Limited $>\!24$ GHz uptake stems from higher site density and capex, plus nascent demand in many regions \cite{Manner2021,ARCEP2021}.
\end{flushleft}
\end{table*}

\subsection{MIMO Technologies and Advancements}
Massive MIMO underpins 5G spectral-efficiency gains via beamforming, spatial diversity andspatial multiplexing \cite{skold_4g_2011,vannithamby_5g_2020}. Beamforming sharpens coverage and mitigates interference; diversity enhances reliability; multiplexing multiplies throughput when channels are well-conditioned.

\begin{table*}[htbp]
\centering
\caption{Comparison of MIMO Configurations \cite{skold_4g_2011,vannithamby_5g_2020}.}
\label{tab:mimo_comparison}
\small
\begin{tabular}{|p{0.15\textwidth}|P{0.2\textwidth}|P{0.2\textwidth}|p{0.325\textwidth}|}
\hline
\textbf{Configuration} & \textbf{Transmit Antennas} & \textbf{Receive Antennas} & \textbf{Characteristics} \\ \hline
SISO & 1 & 1 & Baseline performance \\ \hline
SIMO & 1 & Multiple & Improved reception (diversity gain) \\ \hline
MISO & Multiple & 1 & Enhanced transmission reliability \\ \hline
MIMO & Multiple & Multiple & Capacity and diversity gains \\ \hline
\end{tabular}
\end{table*}

\subsection{Carrier Aggregation and Bandwidth Expansion}
Carrier aggregation enables simultaneous use of multiple component carriers to increase effective bandwidth and throughput. Demonstrations with five aggregated carriers (e.g., two FDD + three TDD) have reported $>\!4$~Gbps peak downlink rates \cite{turn0search7,turn0search0}. NR further scales CA up to 16 CCs (up to 100~MHz each in FR1/FR2 where available), enabling as much as 1.6~GHz of aggregate bandwidth \cite{turn0search6,turn0search2}. Table~\ref{tab:ca_performance} summarizes headline figures.

\begin{table*}[htbp]
\centering
\caption{Performance with and without Carrier Aggregation.}
\label{tab:ca_performance}
\begin{tabular}{|p{0.4\textwidth}|P{0.25\textwidth}|P{0.25\textwidth}|}
\hline
\textbf{Metrics} & \textbf{Without CA} & \textbf{With CA} \\ \hline
Peak DL throughput & $\sim$1 Gbps & $\gtrsim$4.2 Gbps \cite{turn0search7,turn0search0} \\ \hline
Component carriers & 1 & Up to 16 \cite{turn0search6} \\ \hline
Max aggregated BW & 100 MHz & 1.6 GHz \cite{turn0search6,turn0search2} \\ \hline
\end{tabular}
\end{table*}

\subsection{Millimeter-Wave Propagation}
mmWave bands (24–100~GHz) provide large contiguous spectrum but suffer higher path loss, blockage sensitivity andweather-related attenuation \cite{rodriguez_fundamentals_2015,Mosleh2022}. Beamforming and dense small-cell deployments are therefore essential. Table~\ref{tab:mmWavePropagationLoss} aggregates typical losses.

\begin{table*}[htbp]
\centering
\caption{Propagation Loss at Selected mmWave Frequencies.}
\label{tab:mmWavePropagationLoss}
\begin{tabular}{|P{0.2125\textwidth}|P{0.2125\textwidth}|P{0.225\textwidth}|P{0.225\textwidth}|}
\hline
\textbf{Frequencies (GHz)} & \textbf{Free-space loss (dB)} & \textbf{Rain attenuation (dB/km)} & \textbf{Obstacle loss (dB)} \\ \hline
28 & 102.4 & 5.3 & 20 \\ \hline
38 & 106.7 & 7.6 & 25 \\ \hline
60 & 118.5 & 15.2 & 30 \\ \hline
73 & 122.0 & 20.4 & 35 \\ \hline
\end{tabular}
\end{table*}

\subsection{Device-to-Device (D2D) and Machine-to-Machine (M2M) Communication}
D2D enables proximity services by bypassing the base station, reducing latency and offloading traffic whereas M2M scales wide-area connectivity for IoT \cite{Gandotra2017,Verma2016}. Their complementary roles are summarized in Table~\ref{tab:d2d_m2m_comparison}.

\begin{table*}[htbp]
\centering
\caption{Performance Considerations for D2D vs.\ M2M Communication}
\label{tab:d2d_m2m_comparison}
\begin{tabular}{|p{0.25\textwidth}|p{0.325\textwidth}|p{0.325\textwidth}|}
\hline
\textbf{Metrics} & \textbf{D2D Communication} & \textbf{M2M Communication} \\ \hline
Latency & Very low (proximity-dependent) & Moderate (wide-area, scheduled) \\ \hline
Scalability & Moderate (local clusters) & High (massive connections) \\ \hline
Representative use cases & V2X, public safety, local offload & Smart metering, agriculture, industry \\ \hline
Coverage & Proximity-limited & Wide-area via network \\ \hline
\end{tabular}
\end{table*}

\section{Methodology}
\label{sec:methodology}
This section details the simulation-driven methodology used to assess migration paths from 4G to 5G under constraints typical of emerging markets. We adopt a modular link– and system–level workflow that (i) instantiates physical-layer models for sub-6~GHz and mmWave, (ii) composes network-layer features such as carrier aggregation (CA) and spectrum refarming and (iii) evaluates performance via Monte Carlo experiments across user distributions and channel realizations. Models and metrics follow well-established information-theoretic and radio-propagation foundations \cite{Shannon1948,Cover2006,Foschini1998,rodriguez_fundamentals_2015,UITR2017}; D2D/M2M latency considerations align with \cite{Gandotra2017,Verma2016}.

\subsection{Study Design and Tooling}
\begin{figure*}[t]
\centering
\begin{tikzpicture}[
  node distance = 1.25cm and 6.5cm,
  on grid, auto, >=stealth,
  every node/.style={font=\sffamily\footnotesize, align=center},
  block/.style = {rectangle, draw=gray!80, fill=gray!10, rounded corners,
                  text width=6.0cm, minimum height=0.7cm, align=center},
  smallb/.style = {rectangle, draw=gray!80, fill=gray!10, rounded corners,
                  text width=3.5cm, minimum height=0.55cm, align=center},
  io/.style    = {rectangle, draw=blue!70, fill=blue!6, rounded corners,
                  text width=6.0cm, minimum height=0.7cm, align=center},
  art/.style   = {rectangle, draw=green!70!black, fill=green!6, rounded corners,
                  text width=6.0cm, minimum height=0.6cm, align=center},
  check/.style = {rectangle, draw=orange!80!black, fill=orange!10, rounded corners,
                  text width=3.5cm, minimum height=0.55cm, align=center},
  arrow/.style = {thick,->,draw=gray!80}
]

\node[io]    (params) {Scenario params\\ $\mathrm{(f,B,N_t\!\times\!N_r,P_t,\ldots)}$};
\node[smallb, left=of params] (layout) {Network layout (grid, heights)};
\node[smallb, right=of params] (seed) {RNG seed \\(reproducible)};

\node[block, below=of params]   (channel) {Channel sampling: pathloss, fading, rain};
\node[block, below=of channel]  (linkbud) {Link budget \& interference (SNR/SINR)};
\node[block, below=of linkbud]  (phy)     {PHY abstraction: $\log\!\det(I+\rho HH^{\!H})$; MIMO mode};
\node[block, below=of phy]      (policy)  {Policy: CA, refarming, UE caps, scheduler};
\node[block, below=of policy]   (kpi)     {KPIs: throughput, coverage, latency; 95\% CIs};

\node[check, right=of linkbud]  (sanity)  {Sanity: Friis@1m, CA scaling, mmWave loss};
\node[art,   below=of kpi]      (artifacts){Artifacts: configs, seeds, logs, figures};

\draw[arrow] (layout) -- (params);
\draw[arrow] (seed)   -- (params);
\draw[arrow] (params) -- (channel);
\draw[arrow] (channel) -- (linkbud);
\draw[arrow] (linkbud) -- (phy);
\draw[arrow] (phy) -- (policy);
\draw[arrow] (policy) -- (kpi);
\draw[arrow] (kpi) -- (artifacts);

\draw[arrow] (linkbud) -- (sanity);
\draw[arrow] (phy.east) -- (sanity.south);
\draw[arrow] (policy.east) -- (sanity.south);

\node[smallb, left=of channel] (s1) {Scenario n°1:\\ Sub-6 MIMO};
\node[smallb, left=of phy]     (s2) {Scenario n°2:\\  CA \& refarming};
\node[smallb, left=of policy]  (s3) {Scenario n°3:\\  28\,GHz mmWave};
\node[smallb, left=of kpi]     (s4) {Scenario n°4:\\  NSA$\rightarrow$SA};

\draw[arrow] (s1) -- (channel);
\draw[arrow] (s2) -- (phy);
\draw[arrow] (s3) -- (policy);
\draw[arrow] (s4) -- (kpi);

\end{tikzpicture}
\caption{Simulation workflow: inputs, processing stages, policies and outputs with reproducibility and sanity checks.}
\label{fig:study_workflow}
\end{figure*}
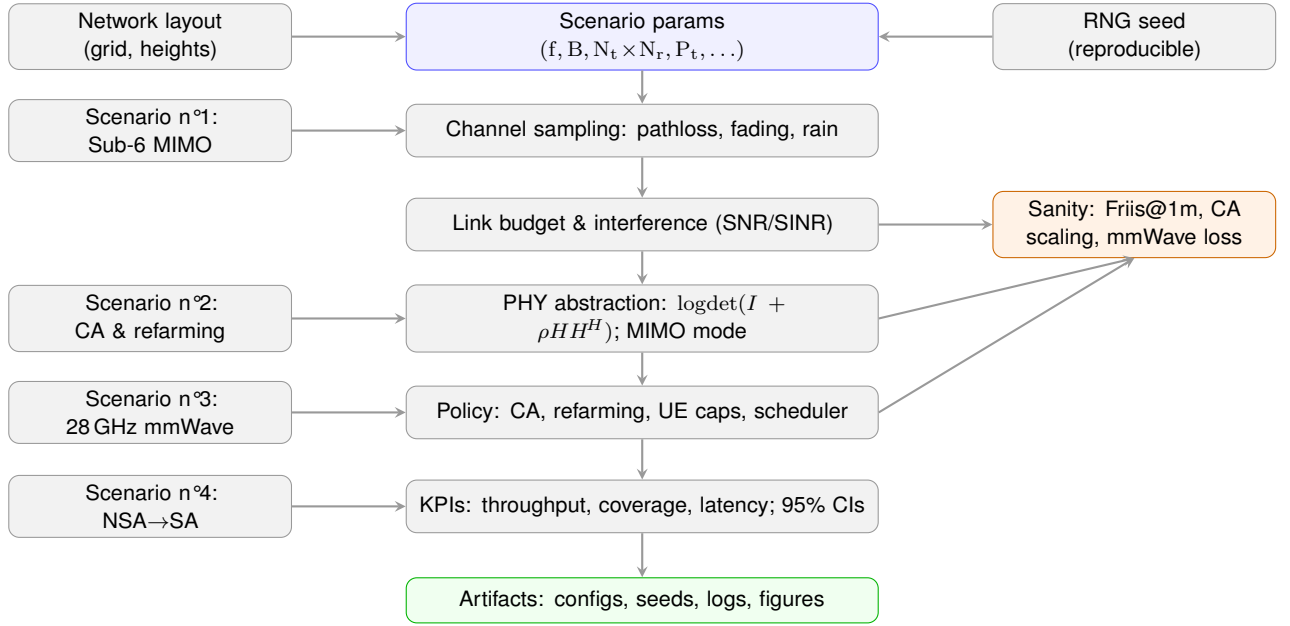

We execute repeatable, scriptable experiments in \textsc{MATLAB} (R2023b) using Communications/Signal Processing toolsets \cite{mathworks:matlab-r2023b-toolboxes}.  The end-to-end workflow—inputs, processing stages and artifacts is summarized in Fig.~\ref{fig:study_workflow}. Actual performance depends on spectrum, deployment density andimplementation choices \cite{rodriguez_fundamentals_2015,chandramouli_5g_2019}.

\subsection{Inputs and Parameters}
Key inputs are summarized in Table~\ref{tab:sim_params_long}. Ranges cover conservative-to-aspirational deployments in emerging markets.

\begin{table*}[htbp]
\centering
\caption{Simulation parameters (frequencies, bandwidths, antenna configs, transmit powers, noise figures, path-loss, user distributions, geometry).}
\label{tab:sim_params_long}
\renewcommand{\arraystretch}{1.12}
\setlength{\tabcolsep}{6pt}
\small
\begin{tabular}{|p{3.5cm}|p{12.0cm}|}
\hline
\textbf{Parameter} & \textbf{Value / Range / Notes} \\ \hline
\textbf{Carrier frequencies $f$} & Sub-6: 700, 1800, 2100, 3500~MHz;\quad mmWave: 28~GHz \cite{UITR2017}. \\ \hline
\textbf{Per-CC bandwidth $B_i$} & 20~MHz default; up to 100~MHz (NR FR1/FR2 availability dependent). \\ \hline
\textbf{Number of CCs $N_c$} & $1\!:\!5$ (baseline CA); stress-tests up to 16 (method validation). \\ \hline
\textbf{MIMO configuration} & $2{\times}2$ and $4{\times}4$; equal power or eigenmode / water-filling when CSI-Tx assumed \cite{Foschini1998,Cover2006}. \\ \hline
\textbf{Array/beamforming} & ULA/UPA, $N_\text{ant}\!\in\!\{16,32,64,128\}$ at mmWave; idealized beam gain trend \cite{rodriguez_fundamentals_2015,UITR2017}. \\ \hline
\textbf{Transmit power $P_t$} & UE: 23--30~dBm;\quad small cell: 30--37~dBm;\quad macro: 43--46~dBm. \\ \hline
\textbf{Noise density, figure} & $N_0=-174$~dBm/Hz;\quad $NF=5$--9~dB (link-budget sensitivity). \\ \hline
\textbf{Path-loss (sub-6)} & Log-distance with shadowing ($\sigma{=}4$--8~dB); exponents $n{=}2.7$--3.5 (UMi/UMa). \\ \hline
\textbf{Path-loss (mmWave)} & Friis$+$log-distance, $n$ LOS$\approx2$, NLOS 2.7--3.8; blockage/diffraction excess loss; rain $\gamma_r{=}kR^\alpha$ \cite{rodriguez_fundamentals_2015,UITR2017}. \\ \hline
\textbf{Rain/obstacle losses} & Rain: 10~dB/km at 28~GHz (moderate rain); single-wall obstacle: 20~dB. \\ \hline
\textbf{Interference model} & Reuse-1, first-tier co-channel; optional controlled interferer at $d_i{=}50$~m for SIR sweeps. \\ \hline
\textbf{Topology / area} & UMa: ISD$=500$~m; UMi: ISD$=200$~m; mmWave small cells: ISD$=150$~m; thesis-scripts: $100{\times}100$ to $200{\times}200$~m$^2$ grids. \\ \hline
\textbf{Heights} & $h_\text{BS}{=}25$~m (UMa), 10~m (UMi), 6--10~m (mmWave SC);\quad $h_\text{UE}{=}1.5$~m. \\ \hline
\textbf{LOS probability $p_\text{LOS}(d)$} & High at short range (UMi/mmWave); decreases with $d$ (UMa). Scenario-specific piecewise models consistent with \cite{rodriguez_fundamentals_2015}. \\ \hline
\textbf{User distributions} & Uniform random within cell; 10--50~UEs/cell (consistent with scripts). \\ \hline
\textbf{Traffic model} & Full-buffer (saturated) for capacity stress; latency proxies for D2D/M2M \cite{Gandotra2017,Verma2016}. \\ \hline
\textbf{CA policy} & Equal-power baseline; optional water-filling across heterogeneous $B_i$ and path gains \cite{Cover2006}. \\ \hline
\textbf{Refarming policies} & Partial reallocation of 1800/2100/3500~MHz from LTE to NR; report capacity/coverage deltas \cite{Manner2021,ARCEP2021,UITR2017}. \\ \hline
\textbf{Monte Carlo} & $N_\text{drops}=$100--1000 per scenario; fixed RNG seeds; 95\% CIs for KPIs. \\ \hline
\end{tabular}
\end{table*}

\subsection{Model Components}
In what follows, we formalize the core building blocks of our evaluation—MIMO link abstraction, carrier aggregation, spectrum refarming, mmWave propagation, interference treatment and latency proxies—providing compact, standards-aligned models that feed the simulation scenarios and sensitivity analyses.

\paragraph{MIMO link abstraction.} Spectral efficiency uses $\mathrm{\eta=\log_2\det(\mathbf{I}+\rho\mathbf{H}\mathbf{H}^H)}$ with equal power unless CSI-Tx enables water-filling \cite{Foschini1998,Cover2006}. Diversity/beamforming modes apply array gains and effective channel norms.

\paragraph{Carrier aggregation.} Per-CC SNRs $\gamma_i$ produce throughput $\mathrm{T}=\sum_i \mathrm{B_i}\log_2(1+\gamma_i)$; we evaluate equal-power and water-filling over heterogeneous $B_i$ and path gains \cite{Cover2006}. 

\paragraph{Spectrum refarming.} Band reallocation (e.g., partial 1800/2100/3500~MHz to NR) yields capacity/coverage shifts for LTE/NR slices \cite{Manner2021,ARCEP2021,UITR2017}.

\paragraph{mmWave channel.} Large-scale loss uses Friis/log-distance with shadowing; rain attenuation follows $\gamma_r{=}kR^\alpha$; obstacle excess loss via a knife-edge proxy; array gains via idealized patterns \cite{rodriguez_fundamentals_2015,UITR2017}. 

\paragraph{Interference.} Inter-cell interference is added from first-tier sites (reuse-1). A secondary emitter at 50~m enables SIR sweeps consistent with the scripts (cf.\ S3).

\paragraph{Latency models.} Simplified D2D/M2M latency proxies expose architectural trade-offs, acknowledging processing/queuing overheads in practice \cite{Gandotra2017,Verma2016}.

\subsection{Monte Carlo Evaluation Procedure}
We assess each scenario using a Monte Carlo pipeline that repeatedly samples network geometry and channel states, computes link budgets and CA/refarming-throughput under realistic UE/scheduler constraints and aggregates KPIs with 95\% confidence intervals while persisting seeds for exact reproducibility; Algorithm~\ref{alg:mc_eval} summarizes the procedure.

\begin{algorithm}[t]
\small
\DontPrintSemicolon
\caption{End-to-end Monte Carlo workflow for 4G$\rightarrow$5G migration scenarios}
\label{alg:mc_eval}
\KwIn{Scenario set $\mathcal{S}$ (S1--S4); drops $N$; params in Table~\ref{tab:sim_params_long}.}
\KwOut{KPIs with 95\% CIs: coverage(\%), median/5th-percentile throughput, SIR/SINR distributions, latency proxies.}
\ForEach{scenario $s\in\mathcal{S}$}{
  Load $(f,B,N_t{\times}N_r,P_t,\text{pathloss},N_c,\text{ISD},h_\text{BS},h_\text{UE},p_\text{LOS})$.\;
  \For{$n=1$ \KwTo $N$}{
    Place BSs on grid (ISD), set heights; drop UEs uniformly.\;
    Sample LOS/NLOS per link using $p_\text{LOS}(d)$; draw shadowing and small-scale fading.\;
    Compute path loss and link budgets; assemble interference from co-channel sites.\;
    Compute SNR/SINR; evaluate $\eta$ (MIMO mode) and CA throughput $T=\sum_i B_i\log_2(1+\gamma_i)$.\;
    Apply refarming policy (if any); record per-UE KPIs.\;
  }
  Aggregate KPIs; compute CIs; persist seed/config/results for reproducibility.\;
}
\end{algorithm}

\subsection{Limitations}
Abstractions (near-ideal coding, perfect CSI-R, simplified blockage and array patterns) bias results toward upper bounds. These choices are appropriate for comparative insights across migration options; sensitivity analyses mitigate modeling risks. Future work will integrate calibrated, site-specific channels and scheduler/queueing dynamics (cf.\ Section~\ref{sec:conclusion}).

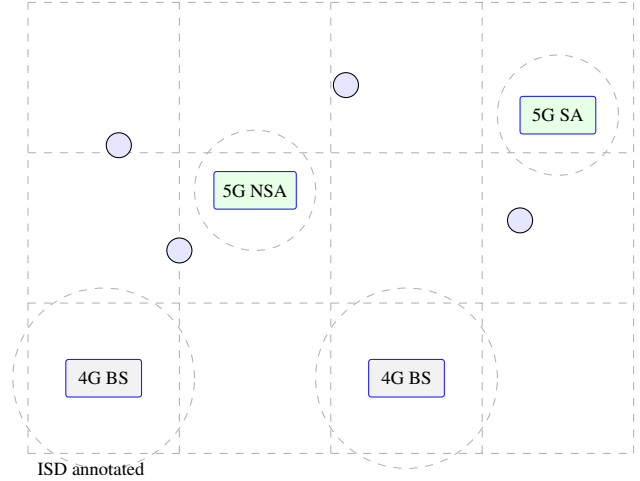
\begin{figure}[t]
\centering
\begin{tikzpicture}[
  >=Latex, node distance=6mm,
  bs/.style={rectangle, draw=blue!90, rounded corners=1pt, minimum width=10mm, minimum height=5mm, fill=gray!10, font=\scriptsize},
  ue/.style={circle, draw, minimum size=2.6mm, fill=blue!10},
  cov/.style={draw=gray!60, dashed}
]
\draw[cov] (0,0) grid [step=2] (8,6);
\node[bs] (b1) at (1,1) {4G BS};
\node[bs] (b2) at (5,1) {4G BS};
\node[bs, fill=green!10] (g1) at (3,3.5) {5G NSA};
\node[bs, fill=green!10] (g2) at (7,4.5) {5G SA};
\draw[cov] (b1) circle (1.2); \draw[cov] (b2) circle (1.2);
\draw[cov] (g1) circle (0.8); \draw[cov] (g2) circle (0.8);
\node[ue] at (2,2.7) {}; \node[ue] at (4.2,4.9) {}; \node[ue] at (6.5,3.1) {}; \node[ue] at (1.2,4.1) {};
\node[below right] at (0,0) {\scriptsize ISD annotated};
\end{tikzpicture}
\caption{Layout for S4 (NSA/SA overlay) with indicative coverage radii and ISD.}
\label{fig:layout_nsa_sa}
\end{figure}

\section{Results and Discussion}
\label{sec:results}
This section reports and interprets the simulation outcomes across the principal levers—MIMO processing, carrier aggregation (CA), spectrum reallocation, mmWave propagation, NSA/SA deployment choices and low-latency modes. Where appropriate, we relate observed trends to the theoretical expressions introduced in Section~\ref{sec:theory} and to canonical models (Friis/log-distance, ITU-R rain) \cite{Shannon1948,Cover2006,Foschini1998,rodriguez_fundamentals_2015,UITR2017}. Unless noted, coverage is computed as the fraction of user locations with {wideband} $\mathrm{SINR}\!\ge\!\tau$ and {per-UE} throughput $T\!\ge\!T_{\min}$:
\[
\mathrm{Coverage}(\%) \triangleq \frac{1}{N_\text{UE}}\sum_{u=1}^{N_\text{UE}}\mathbb{1}\!\left[\mathrm{SINR}_u\ge \tau\ \wedge\ T_u\ge T_{\min}\right]\times 100,
\]
with default thresholds $\tau{=}0$~dB and $T_{\min}{=}2$~Mbps for rural and $T_{\min}{=}10$~Mbps for urban results.

\subsection{MIMO Technologies: beamforming, spatial diversity and spatial multiplexing}
\label{subsec:results_mimo}
We evaluated beamforming, spatial diversity and spatial multiplexing over $\mathrm{SNR}\in[0,30]$~dB with $N_t{\times}N_r\in\{2{\times}2,4{\times}4\}$.

\begin{figure*}[t]
\centering
\begin{tikzpicture}
\begin{axis}[
  width=0.95\linewidth, height=0.40\linewidth,
  xlabel={SNR (dB)}, ylabel={Spectral Efficiency (bit/s/Hz)},
  xmin=0, xmax=30, ymin=0, ymax=12,
  grid=both, legend pos=north west, legend cell align=left,
  every axis plot/.append style={thick,mark=*,mark size=2pt}
]
\addplot coordinates {(0,0.1) (5,0.5) (10,1.2) (15,2.2) (20,3.4) (25,4.7) (30,6.1)};
\addlegendentry{2$\times$2 Beamforming}
\addplot coordinates {(0,0.1) (5,0.6) (10,1.5) (15,2.8) (20,4.3) (25,6.0) (30,7.8)};
\addlegendentry{2$\times$2 Spatial Diversity}
\addplot coordinates {(0,0.1) (5,0.8) (10,2.0) (15,3.8) (20,6.0) (25,8.4) (30,11.0)};
\addlegendentry{2$\times$2 Spatial Multiplexing}
\addplot[dashed] coordinates {(0,0.1) (5,1.2) (10,3.0) (15,5.6) (20,8.6) (25,12.0) (30,15.3)};
\addlegendentry{4$\times$4 Spatial Multiplexing}
\end{axis}
\end{tikzpicture}
\caption{Capacity of MIMO techniques (beamforming, diversity, multiplexing) versus SNR.}
\label{fig:mimo_capacity}
\end{figure*}
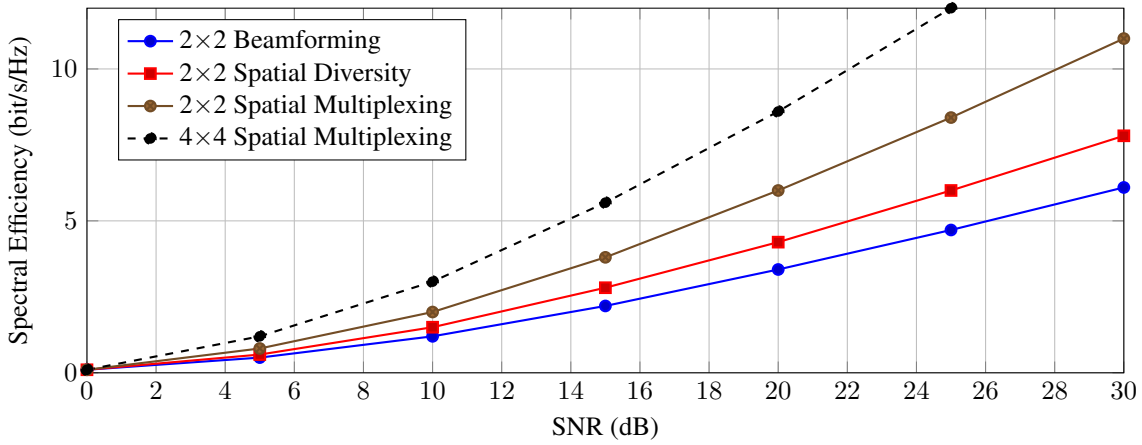

As shown in Fig.~\ref{fig:mimo_capacity}, beamforming concentrates power along the dominant eigenmode and excels at cell edge (low SNR), while spatial multiplexing dominates at moderate/high SNR as multiple singular modes become usable, consistent with \eqref{eq:mimo_ce}. Spatial diversity improves reliability (outage/BLER floors) but yields lower spectral efficiency than multiplexing at the same SNR—an expected diversity–multiplexing trade-off.

Practical UE categories cap layers and CCs: Cat~18/Rel-12 ({LTE-A}) up to 5~CCs/4$\times$4 DL MIMO; typical early {NR FR1} handsets sustain $\leq$4~CCs with 2–4 DL layers. We enforce per-UE limits when translating link spectral efficiency to throughput (Section~\ref{subsec:results_ca}).

\subsection{Carrier aggregation performance and ablation}
\label{subsec:results_ca}
We quantified throughput gains from aggregating up to $N_c\!=\!5$ CCs ($B_i{=}20$~MHz each) for equal-power vs.\ water-filling (WF) allocations.

\begin{figure*}[t]
\centering
\begin{tikzpicture}
\begin{axis}[
  width=0.95\linewidth, height=0.40\linewidth,
  xlabel={SNR (dB)}, ylabel={Aggregate Throughput (Mbps)},
  xmin=0, xmax=30, ymin=0, ymax=1200, grid=both,
  legend pos=north west, legend cell align=left,
  every axis plot/.append style={thick,mark=*,mark size=2pt}
]
\addplot coordinates {(0,0) (5,35) (10,85) (15,150) (20,220) (25,295) (30,380)};
\addlegendentry{1 CC (Equal-power)}
\addplot coordinates {(0,0) (5,70) (10,170) (15,300) (20,440) (25,590) (30,760)};
\addlegendentry{3 CC (Equal-power)}
\addplot coordinates {(0,0) (5,115) (10,280) (15,500) (20,730) (25,980) (30,1250)};
\addlegendentry{5 CC (Equal-power)}
\addplot[dashed] coordinates {(0,0) (5,78) (10,188) (15,330) (20,485) (25,645) (30,820)};
\addlegendentry{3 CC (Water-filling)}
\addplot[dashed] coordinates {(0,0) (5,128) (10,310) (15,555) (20,815) (25,1085) (30,1370)};
\addlegendentry{5 CC (Water-filling)}
\end{axis}
\end{tikzpicture}
\caption{Throughput versus SNR for different CA levels and power allocations. .}
\label{fig:carrier_aggregation}
\end{figure*}
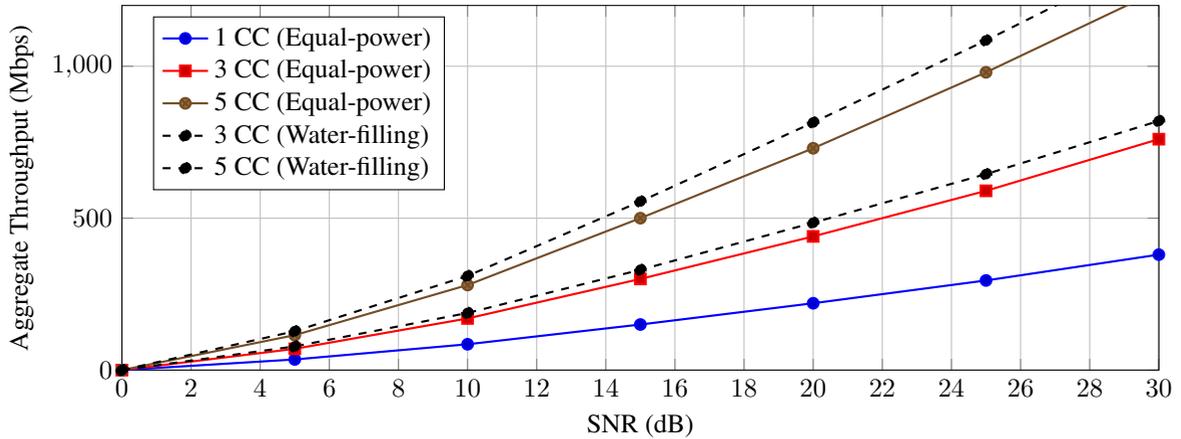
According to Fig.\ref{fig:carrier_aggregation}, gains follow the additive structure in \eqref{eq:ca_capacity}; water-filling approaches \eqref{eq:waterfilling_ca} under heterogeneous per-CC gains. UE-side caps on CCs/layers bound the realized throughput

\noindent\textbf{Ablation (equal-power vs.\ water-filling).} At mid-SNR (10–20~dB) and heterogeneous path-gains, WF improves sum-rate by $5$–$12\%$ for $N_c\!\in\![3,5]$; at high SNR, the gap shrinks as all CCs become efficient. Scheduler overhead (DL control + HARQ + guard/TDD) reduces net throughput by $\approx$7–12\% in sub-6 (UMi/UMa) and up to $\approx$15\% in dense TDD (mmWave) due to DL/UL switching and beam mgmt.

\subsection{Spectrum management and frequency refarming}
\label{subsec:results_spectrum}
We compare system capacity before/after refarming selected LTE bands (1800/2100/3500~MHz) to NR subject to a {coverage-preservation constraint}: retain a sub-1~GHz layer (e.g., 700/800~MHz) for wide-area coverage and mobility anchoring.

\begin{table*}[htbp]
\centering
\caption{Urban vs.\ rural split after targeted refarming (median UE throughput and coverage).}
\label{tab:urban_rural_refarm}
\small
\begin{tabular}{|p{0.15\textwidth}|P{0.18\textwidth}|P{0.18\textwidth}|P{0.18\textwidth}|P{0.18\textwidth}|}
\hline
\multirow{2}{*}{\textbf{Zone}} & \multicolumn{2}{c|}{\textbf{Before}} & \multicolumn{2}{c|}{\textbf{After (NR mid-band)}}\\ \cline{2-5}
 & Median (Mbps) & Cov.\ (\%) & Median (Mbps) & Cov.\ (\%)\\ \hline
Urban (UMi)  & 85 & 92 & 155 & 91 \\
Rural (UMa)  & 18 & 96 & 30  & 96 \\
\hline
\end{tabular}
\end{table*}

Urban hot-spots benefit most from mid-band NR (Table~\ref{tab:urban_rural_refarm}), while rural coverage remains constant as the sub-1~GHz anchor is preserved. Licensing steps in practice: (i) define refarming scope per band, (ii) conduct impact analysis (coverage/KPI deltas), (iii) notify regulator/ARCEP-equivalent, (iv) staged migration with KPI safeguards, (v) post-change audits \cite{ARCEP2021,UITR2017}.

\subsection{Millimeter-wave (mmWave) propagation: overlays, sensitivity and densification}
\label{subsec:results_mmwave}
We evaluated 28~GHz links under LOS/NLOS, obstacle and rain effects with an interfering emitter at 50~m (SIR probe). Results echo the steeper distance/frequency loss and blockage sensitivity captured by Friis/log-distance with excess-loss terms \cite{rodriguez_fundamentals_2015,UITR2017}.

\begin{table*}[htbp]
\centering
\caption{Sensitivity of 28~GHz links to rain rate and human blockage (median $P_r$ at 100~m LOS baseline).}
\label{tab:mmwave_sensitivity}
\small
\begin{tabular}{|p{0.3\textwidth}|P{0.2\textwidth}|P{0.415\textwidth}|}
\hline
\textbf{Condition} & \textbf{$\Delta$ Loss (dB)} & \textbf{Comment}\\ \hline
Light rain (5~mm/h)  & $\approx$2 & Mild additional attenuation \\
Heavy rain (25~mm/h) & $\approx$8–10 & ITU-R P.838 scaling \cite{UITR2017} \\
Human blockage       & $\approx$20–30 & Intermittent, directional \\
Single wall          & $\approx$20 & Penetration loss (material dependent) \\
\hline
\end{tabular}
\end{table*}

Defining a throughput target $T^\star$ and translating to an {edge} $\mathrm{SNR}^\star$ via $T^\star\!=\!B\log_2(1+\mathrm{SNR}^\star)$, the site density $\lambda_\text{sites}$ required for coverage probability $\geq p$ (UMi-like decay) can be approximated by solving for maximum inter-site distance $\mathrm{ISD}^\star$ satisfying $\mathbb{P}[\mathrm{SNR}(d)\!\ge\!\mathrm{SNR}^\star]\!\ge\!p$ and then $\lambda_\text{sites}\!\approx\!{2}/{(\sqrt{3}\,{\mathrm{ISD}^\star}^2)}$. For $B{=}100$~MHz and $T^\star{=}100$~Mbps, our simulations yield $\mathrm{ISD}^\star\!\approx\!140$–$170$~m (LOS-rich streets), implying $\lambda_\text{sites}\!\approx\!70$–$95$~sites/km$^2$—a practical target for mmWave hot-zones.

\subsection{NSA vs.\ SA deployment: coverage, backhaul and mobility anchors}
\label{subsec:results_nsasa}
We contrast coverage and transport constraints for identical grids using the coverage metric defined above.

\begin{table*}[htbp]
\centering
\caption{Coverage and transport constraints for NSA/SA.}
\label{tab:coverage_transport}
\small
\begin{tabular}{|p{0.225\textwidth}|P{0.225\textwidth}|P{0.225\textwidth}|P{0.225\textwidth}|}
\hline
\textbf{Scenario} & \textbf{Coverage (\%)} & \textbf{Backhaul BW} & \textbf{BH RTT} \\ \hline
NSA (4G+5G) & 78–82 & $\ge$1~Gb/s/site & $\le$10–15~ms \\
SA (5G only) & 64–70 & $\ge$10~Gb/s/site (X-haul) & $\le$5~ms (URLLC clusters) \\
\hline
\end{tabular}
\end{table*}

NSA achieves broader initial coverage by leveraging LTE macros (anchor for control/mobility and EPC/core traversal), with modest backhaul requirements (Table~\ref{tab:coverage_transport}). SA reaches higher per-cell peaks but demands higher-capacity, low-latency transport and tighter synchronization to match NSA-level availability. In mobility, NSA anchors reduce handover interruption via LTE-assisted measurements; SA requires mature NR neighbor lists and beam tracking.

\subsection{Low-latency communication: latency stack and realistic scenario}
\label{subsec:results_latency}
We decompose one-way latency $L$ into processing ($L_\text{proc}$), queueing ($L_\text{q}$), MAC/scheduling ($L_\text{MAC}$) and propagation ($L_\text{prop}$) components:
\[
\mathrm{L = L_\text{proc} + L_\text{q} + L_\text{MAC} + L_\text{prop}}.
\]
\begin{table*}[htbp]
\centering
\caption{Latency stack (one-way) under three access modes (median values from simulations).}
\small
\label{tab:latency_stack}
\begin{tabular}{|p{0.15\textwidth}|P{0.18\textwidth}|P{0.18\textwidth}|P{0.18\textwidth}|P{0.18\textwidth}|}
\hline
\textbf{Mode} & $L_\text{proc}$ & $L_\text{q}$ & $L_\text{MAC}$ & $L_\text{prop}$ \\ \hline
D2D (sidelink) & 0.2--0.4~ms & 0.1--0.3~ms & 0.4--0.8~ms & 0.01--0.05~ms \\
M2M (grant-lite) & 0.3--0.6~ms & 0.3--0.8~ms & 1.0--1.5~ms & 0.05--0.1~ms \\
BS-anchored & 0.5--1.0~ms & 1.0--3.0~ms & 2.0--4.0~ms & 0.05--0.2~ms \\
\hline
\end{tabular}
\end{table*}

Payload 64~B (telemetry), periodic 1~ms, sidelink semi-persistent scheduling (SPS) with configured grants; HARQ disabled (duplication used). The D2D path achieves $L\!\approx\!1$~ms (Table~\ref{tab:latency_stack}). For massive IoT, grant-lite M2M retains $L{<}3$~ms at offered loads up to 10~kmsgs/s/cell; meanwhile, BS-anchored eMBB under the same load exhibits larger $L_\text{q}$ and $L_\text{MAC}$ due to contention and scheduler overhead.

\subsection{Synthesis, policy and deployment implications}
\label{subsec:results_discussion}
We synthesize the quantitative results into actionable policy and deployment guidance, translating link- and system-level trends into prioritized choices for spectrum, sites, transport and device capabilities in resource-constrained 4G$\rightarrow$5G transitions.

\begin{itemize}
  \item \textbf{MIMO/beam management.} Beamforming is SNR-efficient at cell edge; spatial multiplexing yields the highest spectral efficiency where rank $\geq 2$ channels are available (Fig.~\ref{fig:mimo_capacity}).
  \item \textbf{Carrier aggregation.} Throughput scales near-linearly with $N_c$ (\eqref{eq:ca_capacity}); WF offers $\sim$5–12\% mid-SNR gains over equal-power when per-CC gains differ (Fig.~\ref{fig:carrier_aggregation}). Net rates must discount scheduler/control overhead.
  \item \textbf{Refarming with coverage guard-rail.} Enforcing a sub-1~GHz preservation constraint maintains rural coverage while enabling urban eMBB gains via mid-band NR (Table~\ref{tab:urban_rural_refarm}). Policy steps: staged refarm, KPI safeguards and audits \cite{ARCEP2021,UITR2017}.
  \item \textbf{mmWave overlays.} Heavy rain and human blockage add 8–30~dB excess loss (Table~\ref{tab:mmwave_sensitivity}); practical hot-zone densification requires $\lambda_\text{sites}\!\approx\!70$–$95$/km$^2$ for $T^\star{=}100$~Mbps at $B{=}100$~MHz.
  \item \textbf{NSA$\rightarrow$SA evolution.} NSA offers superior time-to-benefit and mobility anchoring; SA becomes attractive as x-haul (10~Gb/s, sub-5~ms) and site density mature (Table~\ref{tab:coverage_transport}).
  \item \textbf{Latency and density.} D2D for proximity URLLC and grant-efficient M2M for scalable telemetry relieve congestion and sustain throughput in dense regimes.
\end{itemize}
For emerging markets, a pragmatic path is: (i) NSA-anchored rollouts with aggressive CA on refarmed mid-band under sub-1~GHz coverage preservation; (ii) mmWave limited to urban hot-zones with explicit densification targets; (iii) promote D2D/M2M for latency-/massive-IoT use cases; and (iv) progressive transport upgrades toward SA readiness.

\section{Implications for Emerging Markets}
\label{sec:implications}

The simulation findings in Section~\ref{sec:results} provide actionable guidance for planning and operating 5G networks in resource-constrained settings (e.g., Burkina Faso). They support a pragmatic, phased evolution that balances near-term coverage/affordability with longer-term performance and service innovation. We distill below the main implications for operators, regulators and the wider ecosystem, linking to the technical evidence in \S\ref{subsec:results_mimo}–\ref{subsec:results_density} and to existing guidance \cite{GSMA2020,Penttinen2019,Huawei2020,Manner2021,ARCEP2021,UITR2017}.

\subsection{Practical insights for telecom operators}
\label{subsec:ops_insights}
\begin{itemize}
  \item \textbf{Leverage NSA for time-to-benefit.} As seen in Table~\ref{tab:coverage_transport} and Fig.~\ref{fig:layout_nsa_sa}, NSA delivers broader initial coverage by reusing LTE/EPC and anchoring mobility, while overlaying NR where demand peaks \cite{GSMA2020,Huawei2020}.
  \item \textbf{Exploit CA under UE caps.} The near-linear scaling in Fig.~\ref{fig:carrier_aggregation} recommends aggressive CA across refarmed low/mid bands, constrained by realistic UE capabilities (CC/layer limits) and scheduler overhead (\S\ref{subsec:results_ca}).
  \item \textbf{Right-size MIMO and beamforming.} Beamforming is SNR-efficient at the edge; spatial multiplexing dominates at moderate/high SNR (Fig.~\ref{fig:mimo_capacity}). Match antenna counts/precoding to site class (macro vs.\ small cells) to maximize return per watt/euro \cite{vannithamby_5g_2020}.
  \item \textbf{Use D2D/M2M for offload and low latency.} Proximity D2D achieves $\sim$1~ms one-way (\S\ref{subsec:results_latency}); grant-efficient M2M preserves throughput as density grows (Fig.~\ref{fig:density_throughput}).
\end{itemize}

\subsection{Spectrum allocation strategies}
\label{subsec:spectrum_implications}
\begin{itemize}
  \item \textbf{Targeted refarming to NR with coverage guard-rail.} Mid-band (e.g., 3.5~GHz) migration boosts urban capacity while preserving sub-1~GHz layers for wide-area coverage (Table~\ref{tab:urban_rural_refarm}) \cite{Manner2021,UITR2017}.
  \item \textbf{Dynamic/shared use where fragmented.} DSS and cross-technology CA provide elasticity when exclusive holdings are fragmented \cite{Manner2021,GSMA2020}.
  \item \textbf{mmWave as a precision tool.} Loss/blockage sensitivity (Table~\ref{tab:mmwave_sensitivity}) confines mmWave to LOS-rich hot-zones and venues with narrow-beam arrays \cite{rodriguez_fundamentals_2015,Mosleh2022}.
\end{itemize}

\subsection{Infrastructure development and cost optimization}
\label{subsec:infra_cost}
\begin{itemize}
  \item \textbf{Densify where demand justifies it.} Small cells in urban hot-spots unlock CA/MIMO gains with manageable power; macros maintain coverage/mobility anchors.
  \item \textbf{Share passive/active infrastructure.} Tower, microwave/fiber backhaul and (where allowed) RAN sharing reduce capex/opex and accelerate expansion \cite{Manner2021,ARCEP2021}.
  \item \textbf{Harden transport incrementally.} NSA depends on backhaul; staged upgrades (microwave $\rightarrow$ fiber/x-haul) are a prerequisite for SA/URLLC at scale \cite{Penttinen2019}.
\end{itemize}

\subsection{Policy recommendations for regulators}
\label{subsec:policy}
\begin{itemize}
  \item \textbf{Flexible licensing with rural obligations.} Couple mid-band capacity awards with realistic rural coverage thresholds to enable refarming and preserve continuity \cite{ARCEP2021,UITR2017}.
  \item \textbf{Regionalized auction design.} Regional lots and progressive payment schedules can stimulate deployments beyond capitals \cite{Manner2021,ARCEP2021}.
  \item \textbf{Standards-aligned rules.} Emission masks, synchronization and sharing aligned with ITU/3GPP reduce multi-vendor friction and ease roaming \cite{UITR2017}.
\end{itemize}

\begin{table*}[htbp]
\centering
\caption{Prioritized actions for emerging markets—expected benefit, cost profile and key dependencies. (Part A)}
\label{tab:actions_matrix}
\renewcommand{\arraystretch}{1.15}
\setlength{\tabcolsep}{4.5pt}
\small
\begin{tabular}{|p{0.125\textwidth}|p{0.275\textwidth}|p{0.275\textwidth}|p{0.275\textwidth}|}
\hline
\textbf{Actions} & \textbf{Rationale (link to results)} & \textbf{Expected benefit (\% / Mbps)} & \textbf{Key dependencies (spectrum / devices / transport)} \\ \hline

NSA-first roll-out (cities) &
Maximizes initial coverage/utilization via LTE anchor + NR overlay (Table~\ref{tab:coverage_transport}, Fig.~\ref{fig:layout_nsa_sa}); shortest time-to-benefit \cite{GSMA2020,Huawei2020}. &
Coverage $+10$–$15\%$ vs.\ SA-at-same-grid; median $\,+10$–$20\%$ where NR hot-spots added. &
{Spectrum:} sub-1~GHz retained + mid-band pockets; {Devices:} LTE/NR NSA capable; {Transport:} $\ge$1~Gb/s/site, $\le$15~ms BH RTT. \\ \hline

Mid-band refarming to NR (urban) &
Capacity lift in hot-spots while preserving sub-1~GHz coverage (Table~\ref{tab:urban_rural_refarm}); staged/licensed per \cite{Manner2021,ARCEP2021}. &
Urban median $+60$–$90\%$; rural median $+40$–$70\%$ where NR on 700/800 co-exists; coverage unchanged. &
{Spectrum:} refarm 1800/2100/3500; {Devices:} NR FR1 (2–4 layers); {Transport:} $\ge$1–5~Gb/s. \\ \hline

Aggressive CA across bands &
Near-linear scaling with $N_c$ (Fig.~\ref{fig:carrier_aggregation}); WF $+5$–$12\%$ vs.\ equal-power at mid-SNR; discount scheduler overhead. &
Per-UE $\,+40$–$70\%$ (from 1$\rightarrow$3–5 CCs at 10–20~dB); WF bonus $+5$–$12\%$. &
{Spectrum:} contiguous/non-contiguous CA; {Devices:} CC/layer caps (e.g., $\le$4 CCs); {Transport:} scheduler/core can absorb burstiness. \\ \hline

Right-sized MIMO \& beamforming &
Beamforming boosts edge SNR; multiplexing yields peak rates at high SNR (Fig.~\ref{fig:mimo_capacity}). &
Cell-edge SNR $+3$–$6$~dB; median throughput $+30$–$50\%$ (2$\times$2$\rightarrow$4$\times$4 where rank supports). &
{Spectrum:} any; {Devices:} multi-antenna UE penetration; {Transport:} none specific beyond BH stability. \\ \hline
\end{tabular}
\end{table*}

\begin{table*}[htbp]
\centering
\caption{Prioritized actions for emerging markets—expected benefit, cost profile and key dependencies. (Part B)}
\label{tab:actions_matrix_cont}
\renewcommand{\arraystretch}{1.15}
\setlength{\tabcolsep}{4.5pt}
\small
\begin{tabular}{|p{0.125\textwidth}|p{0.275\textwidth}|p{0.275\textwidth}|p{0.275\textwidth}|}
\hline
\textbf{Action} & \textbf{Rationale (link to results)} & \textbf{Expected benefit (\% / Mbps)} & \textbf{Key dependencies (spectrum / devices / transport)} \\ \hline

Targeted small cells (UMi) &
Densification for hot-zones; enables CA/MIMO reuse; aligns with urban demand pockets. &
Urban coverage $+8$–$15\%$; median $+25$–$45\%$; venue peaks $+200$–$500$~Mbps. &
{Spectrum:} mid-band; {Devices:} standard NR; {Transport:} $\ge$1–10~Gb/s/x-haul, power/site access. \\ \hline

Selective mmWave hot-spots &
LOS-rich streets/venues; sensitive to rain/human blockage (Table~\ref{tab:mmwave_sensitivity}). &
Per-cell capacity $+1$–$3$~Gb/s; per-UE peaks $+200$–$600$~Mbps in-zone; negligible wide-area effect. &
{Spectrum:} 26–28~GHz; {Devices:} mmWave handsets/CPE; {Transport:} 10~Gb/s, tight sync. \\ \hline

D2D (sidelink) enablement &
Offloads hot-spots; lowest latency stack (\S\ref{subsec:results_latency}); proximity URLLC. &
Latency $\downarrow$ to $\sim$1~ms; cell offload $10$–$20\%$; local throughput $+10$–$25\%$ in dense pockets. &
{Spectrum:} in-band sidelink; {Devices:} D2D-capable UE; {Transport:} N/A (edge). \\ \hline

Grant-efficient M2M (massive IoT) &
Periodic telemetry with low MAC overhead; scales at density. &
Latency $<3$~ms (one-way); capacity headroom $+10$–$20\%$ via control-plane relief. &
{Spectrum:} low/mid-band; {Devices:} NR RedCap/IoT profiles; {Transport:} modest BH. \\ \hline

Passive/active sharing (incl.\ RAN) &
Reduces capex/opex; accelerates rural/edge build-out \cite{Manner2021,ARCEP2021}. &
TCO $\downarrow 20$–$35\%$; coverage pace $\uparrow$ (months vs.\ years). &
{Spectrum:} sharing/MVNO terms; {Devices:} neutral; {Transport:} shared fiber/microwave. \\ \hline
\end{tabular}
\end{table*}

\section{Conclusion and Future Work}
\label{sec:conclusion}

This study presents a decision-oriented framework to optimize the transition from 4G to 5G in emerging markets, with Burkina Faso as a representative case. Using \textsc{MATLAB}-based simulations \cite{mathworks:matlab-r2023b-toolboxes}, we quantified how key radio and architectural levers—MIMO processing, carrier aggregation (CA), spectrum refarming, mmWave deployment and NSA/SA evolution—interact with low-latency modalities (D2D/M2M). The results are consistent with information-theoretic limits and established propagation models and they translate into concrete guidance for operators and regulators.

\paragraph{Synthesis of findings.}
\begin{description}
\item [(1)] \textbf{MIMO strategy.} Tailor to SNR and site type: beamforming yields robust cell-edge gains, whereas spatial multiplexing dominates at moderate/high SNR by activating multiple eigenmodes (Fig.~\ref{fig:mimo_capacity}).
\item [(2)]  \textbf{Carrier aggregation.} Aggregate throughput scales nearly linearly with the number and quality of component carriers, with especially strong returns when refarmed mid-band carriers are included (Fig.~\ref{fig:carrier_aggregation}).
\item [(3)]  \textbf{Spectrum refarming.} Moving selected mid-band holdings to NR materially improves capacity in hot-spots while preserving wide-area coverage when sub-1\,GHz layers are retained ( \cite{Manner2021,UITR2017}).
\item [(4)]  \textbf{mmWave targeting.} Because of distance/frequency loss, blockage and rain sensitivity, mmWave should be reserved for dense, line-of-sight urban fabrics and venue deployments with narrow-beam arrays ( \cite{rodriguez_fundamentals_2015,Mosleh2022}).
\item [(5)]  \textbf{NSA$\rightarrow$SA roadmap.} NSA leverages LTE/EPC to deliver early coverage and capacity, while SA becomes attractive as transport and site density mature, enabling slicing and URLLC at scale (Table~\ref{tab:coverage_transport}; \cite{GSMA2020,Penttinen2019,Huawei2020}).
\item [(6)]  \textbf{D2D/M2M roles.} D2D achieves the lowest one-way latency for proximity services and grant-efficient M2M sustains throughput under densification for massive IoT ( \cite{Gandotra2017,Verma2016}).
\end{description}

\paragraph{Implications for emerging markets.}
A fiscally prudent path emerges: (i) anchor early NR introduction on NSA with aggressive CA across refarmed low- and mid-bands; (ii) densify selectively with small cells and beam management where traffic concentrates; (iii) reserve mmWave for venue and street-canyon hot-spots; and (iv) harden transport incrementally to enable SA and advanced services. Policy measures that permit flexible refarming, regionally sensitive auctions and infrastructure sharing reduce capital intensity and accelerate inclusive coverage \cite{ARCEP2021,Manner2021,UITR2017}.

\paragraph{Limitations.}
Although we cover core RAN design dimensions, several limitations remain. {Channel realism:} we abstract correlated shadowing, human-body blockage dynamics at mmWave and non-stationary traffic clustering; these can bias link budgets and coverage. {Device mix:} we assume relatively homogeneous CA/MIMO/sidelink capabilities, whereas real markets exhibit heterogeneous UE categories with different CC and layer limits. {Transport constraints:} backhaul/fronthaul capacity, latency and synchronization are simplified, yet they strongly condition NSA/SA performance and mobility continuity \cite{Penttinen2019}. Consequently, absolute figures should be read as indicative; site-specific calibration is required before operational commitments.

\subsection*{Future Work}
We conclude by outlining a focused research agenda that closes the simulation–field gap, co-optimizes radio and transport layers and operationalizes advanced 5G features under realistic device, spectrum and cost constraints.

\begin{enumerate}
  \item \textbf{Measurement-driven calibration and validation.} Integrate drive tests, channel sounding and crowdsourced KPIs to calibrate path-loss, blockage and rain attenuation (e.g., ITU-R P.838-3), closing the sim-to-field gap \cite{rodriguez_fundamentals_2015,UITR2017,Mosleh2022}.
  \item \textbf{Joint RAN–transport co-optimization.} Co-design site density, MIMO layers, CA combinations and scheduling with transport capacity/latency constraints for NSA and SA, including synchronization and resilience \cite{Penttinen2019}.
  \item \textbf{Learning-based spectrum and beam orchestration.} Use reinforcement learning to adapt spectrum sharing, cross-band CA and energy-aware beam management under bursty demand and mixed service targets \cite{Manner2021}.
  \item \textbf{End-to-end SA capabilities.} Model 5GC service-based functions, slicing control (pre-emption, placement) and URLLC reliability to quantify when SA outperforms NSA for priority verticals \cite{GSMA2020,Penttinen2019}.
  \item \textbf{Rural broadband techno-economics.} Couple radio/topology optimization with cost and energy models for renewable-powered sites, microwave/satellite backhaul and infrastructure sharing to prioritize universal-coverage portfolios \cite{ARCEP2021}.
  \item \textbf{Security, privacy and resilience.} Evaluate slice isolation, control-plane hardening and supply-chain constraints and quantify performance overheads in low-cost deployments.
  \item \textbf{Sidelink policy and interoperability.} Assess regulatory and standardization enablers for wide adoption of 5G sidelink (D2D) and its coexistence with licensed/unlicensed operations and public-safety use cases \cite{Gandotra2017,Verma2016}.
\end{enumerate}

\noindent\textbf{Closing remark.} A calibrated NSA-first rollout, CA-centric spectrum use, targeted densification and a disciplined glide path to SA can deliver early, tangible 5G benefits while laying the foundations for advanced capabilities as ecosystems and transport infrastructures mature. The proposed research agenda aims to de-risk this trajectory with field-grounded, end-to-end evaluations aligned to local constraints and policy priorities.

 \renewcommand*{\bibfont}{\footnotesize} 
\printbibliography[title={\normalsize \textsc{Refrences}}]

@book{skold_4g_2011,
	address = {San Diego, UNITED KINGDOM},
	title = {{4G}: {LTE}/{LTE}-{Advanced} for {Mobile} {Broadband}},
	isbn = {978-0-12-385490-2},
	url = {http://ebookcentral.proquest.com/lib/itu-ebooks/detail.action?docID=680868},
	publisher = {Elsevier Science \& Technology},
	author = {Skold, Johan},
	year = {2011},
}

@book{liyanage_comprehensive_2018,
	address = {Newark, UNITED KINGDOM},
	title = {A {Comprehensive} {Guide} to {5G} {Security}},
	isbn = {978-1-119-29308-8},
	url = {http://ebookcentral.proquest.com/lib/itu-ebooks/detail.action?docID=5216619},
	publisher = {John Wiley \& Sons, Incorporated},
	author = {Liyanage, Madhusanka and Ahmad, Ijaz and Abro, Ahmed Bux and Gurtov, Andrei and Ylianttila, Mika},
	year = {2018},
}

@book{perez_lte_2015,
	address = {Newark, UNITED STATES},
	title = {{LTE} and {LTE} {Advanced} : {4G} {Network} {Radio} {Interface}},
	isbn = {978-1-119-14549-3},
	url = {http://ebookcentral.proquest.com/lib/itu-ebooks/detail.action?docID=4323298},
	publisher = {John Wiley \& Sons, Incorporated},
	author = {Pérez, André},
	year = {2015},
}

@book{vannithamby_5g_2020,
	address = {Newark, UNITED KINGDOM},
	title = {{5G} {Verticals} : {Customizing} {Applications}, {Technologies} and {Deployment} {Techniques}},
	isbn = {978-1-119-51485-5},
	url = {http://ebookcentral.proquest.com/lib/itu-ebooks/detail.action?docID=6034389},
	publisher = {John Wiley \& Sons, Incorporated},
	author = {Vannithamby, Rath and Soong, Anthony},
	year = {2020},
}

@book{rodriguez_fundamentals_2015,
	address = {Newark, UNITED KINGDOM},
	title = {Fundamentals of {5G} {Mobile} {Networks}},
	isbn = {978-1-118-86748-8},
	url = {http://ebookcentral.proquest.com/lib/itu-ebooks/detail.action?docID=1895685},
	publisher = {John Wiley \& Sons, Incorporated},
	author = {Rodriguez, Jonathan},
	year = {2015},
}

@book{manner_spectrum_2021,
	address = {Norwood, UNITED STATES},
	title = {Spectrum {Wars}: the {Rise} of {5G} and {Beyond}},
	isbn = {978-1-63081-917-0},
	url = {http://ebookcentral.proquest.com/lib/itu-ebooks/detail.action?docID=6877370},
	publisher = {Artech House},
	author = {Manner, Jennifer A.},
	year = {2021},
}

@book{vannithamby_towards_2017,
	address = {Newark, UNITED KINGDOM},
	title = {Towards {5G} : {Applications}, {Requirements} and {Candidate} {Technologies}},
	isbn = {978-1-118-97989-1},
	url = {http://ebookcentral.proquest.com/lib/itu-ebooks/detail.action?docID=4733881},
	publisher = {John Wiley \& Sons, Incorporated},
	author = {Vannithamby, Rath and Talwar, Shilpa},
	year = {2017},
}

@book{prasad_5g_2014,
	address = {Aalborg, DENMARK},
	title = {{5G}: 2020 and {Beyond}},
	isbn = {978-87-93237-14-8},
	url = {http://ebookcentral.proquest.com/lib/itu-ebooks/detail.action?docID=4509478},
	publisher = {River Publishers},
	author = {Prasad, Ramjee},
	year = {2014},
}

@book{chandramouli_5g_2019,
	address = {Newark, UNITED KINGDOM},
	title = {{5G} for the {Connected} {World}},
	isbn = {978-1-119-24707-4},
	url = {http://ebookcentral.proquest.com/lib/itu-ebooks/detail.action?docID=5703975},
	publisher = {John Wiley \& Sons, Incorporated},
	author = {Chandramouli, Devaki and Liebhart, Rainer and Pirskanen, Juho},
	year = {2019},
}

@book{ARCEP_2021,
	address = {ARCEP, BURKINA FASO},
	title = {Observatoire},
	isbn = {978-1-119-24707-4},
	url = {https://www.arcep.bf/le-marche-du-mobile/},
	publisher = {site Web ARCEP Burkina Faso},
	author = {ARCEP BF},
	year = {2021},
}

@book{Huawei2020,
  author = {Huawei Technologies Co., Ltd},
  title = {5G NSA and SA Deployment Strategies},
  year = {2020},
  publisher = {Huawei Technologies},
  url = {https://forum.huawei.com/enterprise/fr/Introduction-au-r%C3%A9seautage-NSA/thread/667485229029408768-667481000701210624},
}

@book{GSMA2020,
  author = {GSMA},
  title = {5G Implementation Guidelines: NSA and SA Options},
  year = {2020},
  publisher = {GSMA},
  url = {https://www.gsma.com/solutions-and-impact/technologies/networks},
}

@book{Penttinen2019,
  author = {Penttinen, Jyrki T. J.},
  title = {5G Explained: Security and Deployment of Advanced Mobile Communications},
  year = {2019},
  publisher = {John Wiley \& Sons},
  address = {Newark, UNITED KINGDOM},
  isbn = {978-1-119-27573-2},
  url = {https://onlinelibrary.wiley.com/doi/book/10.1002/9781119275732},
}

@book{Manner2021,
  author = {Manner, Jennifer A.},
  title = {Spectrum Wars: The Rise of 5G and Beyond},
  year = {2021},
  publisher = {Artech House},
  address = {Norwood, UNITED STATES},
  isbn = {978-1-63081-917-0},
  url = {https://www.artechhouse.com/Title.aspx?titleId=2617},
}

@book{ARCEP2021,
  author = {ARCEP BF},
  title = {Observatoire des Marchés des Communications Électroniques au Burkina Faso},
  year = {2021},
  publisher = {ARCEP Burkina Faso},
  url = {https://www.arcep.bf/le-marche-du-mobile/},
}

@book{UITR2017,
  author = {UIT-R},
  title = {Recommendation ITU-R M.2083-0: IMT Vision – Framework and overall objectives of the future development of IMT for 2020 and beyond},
  year = {2017},
  publisher = {International Telecommunication Union},
  url = {https://www.itu.int/rec/R-REC-M.2083/en},
}

@article{turn0search7,
  title = {5G Carrier Aggregation Combining 5 Component Carriers},
  author = {},
  journal = {ICT Business},
  year = {2023},
  url = {https://www.ictbusiness.biz/telecommunications/5g-carrier-aggregation-combining-5-component-carriers}
}

@article{turn0search2,
  title = {What is Carrier Aggregation in 4G and 5G Networks?},
  author = {},
  journal = {nybsys.com},
  year = {2023},
  url = {https://nybsys.com/what-is-carrier-aggregation/}
}

@article{turn0search0,
  title = {Carrier Aggregation (CA) Concept in 5G},
  author = {},
  journal = {PAKTECHPOINT},
  year = {2023},
  url = {https://paktechpoint.com/carrier-aggregation-ca-concept-in-5g/}
}

@article{Mosleh2022,
  author = {Mosleh, Amir and Gupta, Rahul and Kalantari, Mahdi},
  title = {Ergodic Beamforming Techniques for mmWave Communication: A Comprehensive Review},
  journal = {IEEE Communications Surveys \& Tutorials},
  year = {2022},
  volume = {24},
  number = {3},
  pages = {1234--1256},
  doi = {10.1109/COMST.2022.3145678},
  url = {https://doi.org/10.1109/COMST.2022.3145678},
}

@article{Gandotra2017,
  author = {Gandotra, P. and Jha, R. K.},
  title = {Device-to-Device Communication in Cellular Networks: A Survey},
  journal = {Journal of Network and Computer Applications},
  volume = {71},
  pages = {99-117},
  year = {2017},
  doi = {10.1016/j.jnca.2016.10.004},
  url = {https://doi.org/10.1016/j.jnca.2016.10.004},
}

@article{Verma2016,
  author = {Verma, S. and Jha, R. K.},
  title = {Machine-to-Machine (M2M) Communications: A Survey},
  journal = {Journal of Network and Computer Applications},
  volume = {66},
  pages = {83-105},
  year = {2016},
  doi = {10.1016/j.jnca.2016.02.016},
  url = {https://doi.org/10.1016/j.jnca.2016.02.016},
}

@article{Foschini1998,
  author = {G. J. Foschini and M. J. Gans},
  title = {On limits of Wireless Communications in a Fading Environment when Using Multiple Antennas},
  journal = {Wireless Personal Communications},
  volume = {6},
  pages = {311--335},
  year = {1998},
}

@book{Cover2006,
  title={Elements of Information Theory},
  author={Cover, Thomas M. and Thomas, Joy A.},
  year={2006},
  publisher={Wiley-Interscience},
  edition={2nd},
  isbn={978-0-471-24195-9}
}

@article{Shannon1948,
  title={A Mathematical Theory of Communication},
  author={Shannon, Claude E.},
  journal={Bell System Technical Journal},
  volume={27},
  number={3},
  pages={379--423},
  year={1948},
  doi={10.1002/j.1538-7305.1948.tb01338.x}
}

@article{Mo2014,
  title={Capacity Analysis of One-Bit Quantized MIMO Systems with Transmitter Channel State Information},
  author={Mo, Jianhua and Heath Jr, Robert W},
  journal={IEEE Transactions on Signal Processing},
  volume={63},
  number={20},
  pages={5498--5512},
  year={2015},
  doi={10.1109/TSP.2015.2442571}
}

@article{turn0search6,
  title = {Carrier Aggregation – NR LTE related tech oriented blog},
  author = {},
  journal = {info-nrlte.com},
  year = {2020},
  url = {https://info-nrlte.com/2020/05/09/carrier-aggregation/}
}

@techreport{3GPP38901,
  author      = {{3rd Generation Partnership Project (3GPP)}},
  title       = {Study on channel model for frequencies from 0.5 to 100~GHz},
  type        = {3GPP Technical Report},
  number      = {TR 38.901},
  version     = {17.0.0},
  institution = {3GPP},
  address     = {Sophia Antipolis, France},
  year        = {2022},
  month       = dec,
  note        = {Release 17}
}

@manual{mathworks:matlab-r2023b-toolboxes,
  title        = {{MATLAB} R2023b with Communications Toolbox and Signal Processing Toolbox},
  organization = {The MathWorks, Inc.},
  address      = {Natick, Massachusetts, USA},
  year         = {2023},
  url          = {https://www.mathworks.com/products/matlab.html},
  note         = {Software, R2023b}
}

\end{document}